\documentclass[3p,twocolumn]{elsarticle}
\usepackage{amsmath,amssymb,amsfonts}
\usepackage{algorithmic}
\usepackage{algorithm}
\usepackage{graphicx}
\usepackage{textcomp}
\usepackage{tabularx}
\usepackage{caption}

\usepackage{booktabs}

\usepackage{subcaption}

\usepackage{changepage} 

\usepackage{siunitx}

\usepackage{lineno}

\usepackage{mdframed} 

\usepackage{hyperref}

\usepackage{mathtools}

\usepackage{color}
\usepackage{tikz}
\usepackage{pgfplots}
\pgfplotsset{compat=1.12}
\usetikzlibrary{patterns}
\usepackage{pgfplotstable}
\usepackage{csvsimple}
\definecolor{excelblue}{RGB}{94,156,211}
\definecolor{excelorange}{RGB}{235,125,60}
\definecolor{excelgray}{RGB}{165,165,165}
\definecolor{excelgreen}{RGB}{114,172,77}

\usepackage{xcolor,colortbl}
\definecolor{Gray}{gray}{0.9}
\newcolumntype{G}{>{\columncolor{Gray}}r}

\definecolor{tablecolor1}{HTML}{F0F0F0}
\definecolor{tablecolor2}{HTML}{FFFFFF}

\definecolor{mycolor0}{HTML}{2B83BA}
\definecolor{mycolor1}{HTML}{D7301F}
\definecolor{mycolor2}{HTML}{FC8D59}
\definecolor{mycolor3}{HTML}{35A34A}
\definecolor{mycolor4}{HTML}{AF5CBC}
\definecolor{mycolor5}{HTML}{FEF0D9}

\makeatletter
\newcommand{\leqnos}{\tagsleft@true\let\veqno\@@leqno}
\newcommand{\reqnos}{\tagsleft@false\let\veqno\@@eqno}
\reqnos
\makeatother

\def\bfa{{\mathbf a}}

\def\bfq{{\mathbf q}}

\def\bft{{\mathbf t}}

\def\bfx{{\discr{x}}}

\def\bfmu{{\discr{\mu}}}

\def\Q{{\discr{Q}}}

\def\Y{{\discr{Y}}}

\def\C{{\mathcal{C}}}
\def\N{{\mathcal{N}}}

\def\0{{\mathbf 0}}

\newcommand \discr[1]{{\boldsymbol{#1}}}   

\def\superstar{^{\raise 0.5pt\hbox{$\nthinsp *$}}}

\def\nthinsp{\mskip -2   mu}
\makeatletter
\newcommand*{\transpose}{%
	{\mathpalette\@transpose{}}%
}
\newcommand*{\@transpose}[2]{%
	\raisebox{\depth}{$\m@th#1\intercal$}%
}
\makeatother

\usepackage{lineno,hyperref}
\nolinenumbers

\journal{Journal of Computational Science}

\begin{document}
	
	\begin{frontmatter}
		
		\title{Sensitivity Analysis of High-Dimensional Models with Correlated Inputs\tnoteref{mytitlenote}}
		\tnotetext[mytitlenote]{This research is part of the activities of the Innosuisse project no 34394.1 entitled “High-Performance Data Analytics Framework for Power Markets Simulation”
				which is financially supported by the Swiss Innovation Agency.
				This work was supported by a grant from the Swiss National Supercomputing Centre (CSCS) under project ID~d120.
				DS and DG have been supported by the SEAVEA ExCALIBUR project, which has received funding from EPSRC under grant agreement EP/W007711/1.
			}
		
		\author[USI]{Juraj~Kardo\v s\corref{mycorrespondingauthor}}
		\cortext[mycorrespondingauthor]{Corresponding author}
		\ead{juraj.kardos@usi.ch}
		
		\author[CWI]{Wouter~Edeling}

		\author[Brunel]{Diana~Suleimenova}
		\author[Brunel]{Derek~Groen}
		\author[USI]{Olaf~Schenk}

		\address[USI]{Institute of Computing, Universit\`a della Svizzera italiana, Via Buffi 13, 6900 Lugano, Switzerland}
		\address[CWI]{Centrum Wiskunde \& Informatica, Science Park 123, 1098 XG Amsterdam, Netherlands}
		\address[Brunel]{Department of Computer Science, Brunel University London, Kingston Lane,	Uxbridge,	Middlesex UB8 3PH, UK}

		\begin{abstract}
   Sensitivity analysis is an important tool used in many domains of computational science to either gain insight into the mathematical model and interaction of its parameters or study the uncertainty propagation through the input-output interactions. In many applications, the inputs are stochastically dependent, which violates one of the essential assumptions in the state-of-the-art sensitivity analysis methods. Consequently, the results obtained ignoring the correlations provide values which do not reflect the true contributions of the input parameters. 
    This study proposes an approach to address the parameter correlations using a polynomial chaos expansion method and Rosenblatt and Cholesky transformations to reflect the parameter dependencies.
    Treatment of the correlated variables is discussed in context of variance and derivative-based sensitivity analysis.
	We demonstrate that the sensitivity of the correlated parameters can not only differ in magnitude, but even the sign of the derivative-based index can be inverted, thus significantly altering the model behavior compared to the prediction of the analysis disregarding the correlations. 
    Numerous experiments are conducted using workflow automation tools within the VECMA toolkit.
\end{abstract}
		
		
	\end{frontmatter}
	
	
	\section{Nomenclature}


\begin{center}
	\begin{tabular}{ m{2em} m{17em} } 
	$\Q$ & A set of uncertain input parameters $Q_i$ \\
	$D$ & A number of the input parameters \\
	$\rho_{Q_i}$ & Parameter probability density function\\
	$\bfq$ & A set of parameter realizations \\
	$\Y$ & Vector of the application model outputs \\
	$U$ & An application model $\Y = U(\bft,\bfx,\Q)$ \\
	$P$ & Degree of the polynomial  basis \\
	$\Psi$ & Polynomial basis\\
	$\bfa$ & Polynomial coefficients for the basis $\Psi$\\
	$\hat{(\cdot)}$ & Quantities related to the polynomial approximation of the true model \\
	\end{tabular}
\end{center}

\begin{center}
	\begin{tabular}{ m{2.5em} m{19em} } 
		$\mathbb{V}$ & Variance operator \\ 
		$\mathbb{E}$ & Expectation value operator \\ 
		$S_i$ & Variance-based sensitivity index\\
		$S_i^\mathcal{D}$ & Derivative-based  sensitivity index\\
		%
		$(\cdot)^*$ & Denotes the correlated variables/samples \\
		$\bfmu$ & Mean vector of the uncertain parameters \\
		$\mathcal{C}$ & Covariance matrix of the parameters\\
		$C$ & Correlation matrix  of the parameters\\
		$L$ & Cholesky factor of the correlation matrix \\
		$\mathcal{P}$ & Permutation vector \\
		$\kappa$ & Dissipation rate of the container \\
		$T_{env}$ & Ambient temperature \\
		$T_0$	 & Initial temperature of the liquid
	\end{tabular}
\end{center}


	\section{Introduction}
Sensitivity analysis (SA) is a technique for understanding how changes in the input parameters influence the uncertainty in the output of a model or simulation. SA facilitates the understanding of how the outputs of a model change with respect to variations in the input parameters. It  it particularly useful for complex models, in order to determine which parameters cause the greatest variation of the output, and quantify the sensitivity of the model to changes in these parameters. Additionally, SA can be used to improve the accuracy of a model by identifying and reducing sources of uncertainty in the input data.
Two SA methods are studied in this manuscript, a global variance-based method, where  the sensitivity is computed over the support of the input distributions. A local derivative-based method is considered as well, where the sensitivity is studied only in the vicinity of a fixed input point.



The variance-based SA method~\cite{saltelli2004sensitivity}
quantifies the sensitivity of each input parameter by estimating its contribution to the overall variance of the model output.
This is achieved by decomposing the  variance of the model output by splitting it into contributions which arise due to the impact of the input parameters or their interaction, and the parameters are assigned a sensitivity index based on their relative contributions.
This sensitivity index is also known as the Sobol index~\cite{SOBOL2001}. Variance-based methods allow full exploration of the input space, accounting also for the interactions and nonlinear responses.
The variance-based sensitivity is used especially in the context of uncertainty quantification, where the input parameters are usually characterized by a probability density function, modeling their uncertain nature or reflecting the uncertainty in the data collection method. 
%
The current state-of-the-art of variance-based SA comprises two main methodologies - quasi-Monte Carlo (QMC)~\cite{SOBOL2001} and the methods based on model surrogates such as polynomial chaos expansion (PCE)~\cite{chaospy,uncertainpy}. Both approaches are based on sampling the input parameters from the given probability distributions, where the model is evaluated for the values of the parameter samples. In case of the QMC approach, this process is repeated thousands of times, and statistical metrics such as the mean and variance are computed from the resulting series of model outputs. On the other hand, the general idea behind PCE is to approximate the model input–output relationship  with a polynomial expression, which is then used to directly obtain the statistical metrics such as mean and variance, while the first and total-order Sobol indices can also be calculated directly from the polynomial model~\cite{SUDRET2008}.

In case of the derivative-based analysis, the sensitivity information comprise computation of the partial derivative of the model output with respect to an input parameter at some fixed point in the input space. 
The analytical derivative is often unknown, thus standard methods such as finite differences (FD) are used. The domain of the FD study is local, since such analysis can consider only vicinity around a single parameter and its fixed operating point. However, this shortcoming  can be circumvented by exploiting a surrogate model  where the derivative can be computed analytically. This, in turn,  allows one to study sensitivity considering interactions between multiple parameters via correlations. In this manuscript, the derivative-based sensitivity indices are computed from the PCE surrogate model, in order to obtain information about the interaction of correlated variables.

While the variance-based SA is used more during the initial phases of model design, where the goal is to understand the behavior of a model or simulation and the sources of uncertainty in its inputs. It can be used to guide model calibration by identifying the most important parameters, determine the range of input values that result in acceptable output values. Similarly, it can be used to optimize the design of a system by identifying the inputs that have the greatest impact on the performance of the system, and exploring the trade-offs between different design options~\cite{VECMAapps}. On the other hand, the derivative-based sensitivity is particularly useful in operational context, where it is used to understand how changes in the input variables affect the output of a system or process. This can guide design of the robust control systems which are resilient to variations in the system's inputs. Alternatively, it can be used to manage risk by understanding how changes in input variables affect the risk of a system or process. For example, in finance, partial derivatives can be used to calculate the sensitivity of the value of a portfolio to changes in the underlying asset prices or guide forward hedging ratios in commodity trading~\cite{HullJohn}. 

\subsection{Motivation and Research Context}
Correlation of the input parameters is a common phenomenon in many scientific and engineering models and there have been few studies conducted on sensitivity analyses with correlated parameters. 
Since the standard SA methods assume that the parameters are stochastically independent, this can have a significant impact on the results of the analysis. The presence of parameter correlations renders several assumptions no longer valid, e.g. the polynomials in the PCE are no longer orthogonal.
Additionally, if two input parameters are highly correlated, it may be erroneous to draw conclusions about which of the two parameters has a greater impact on the output of the model using standard SA methods. Similarly, the results do not provide adequate information to determine the sensitivity of the model to variations in these inputs. Consequently, the presence of correlation between the input parameters can lead to biased estimates of the model sensitivity, which can lead to incorrect conclusions about the importance of the inputs and the input-output interactions.
For example, when considering the context of energy market models, the input parameters such as the cost of fossil fuel resources (liquid fuels and natural gas) account for the majority of the variance in the total energy system cost.  However, these parameters are often tightly correlated, and applying the state-of-the-art SA methods ignoring the correlation may lead to an optimistic risk assessment of voltage instability, the cost of power generation, a line overload risk, and a power shortage expectation~\cite{SHUAI2021106916,KARDOS2022108613}.
\subsection{Literature Review and Related Work}
There are two directions in the literature how to deal with the correlations during SA; (i) decomposition of the traditional sensitivity indicies into correlated and uncorrelated parts~\cite{Caniou2012,LI2010} and (ii) introducing new sets of indices which contain all correlations and indices which are reduced by the contributions due to the correlation~\cite{MARA2021,MARA2012,MARA2015}.

The definition of the first order Sobol indices was extended to consider parameter dependencies in~\cite{KUCHERENKO2012}. The method extends the QMC framework, such that the sampling is performed considering the conditional probability densities of the individual inputs. In case of dependent normal distributions, the samples are transformed using the Cholesky decomposition of the correlation matrix. The number of model evaluations required to obtain both the first and total order indices for a simple linear model with three inputs was $2^{16}$, which is prohibitive for real-world complex models.  The interpretation of the indices is also not clear, as in some cases the total Sobol index is smaller than the first order one.

In~\cite{Caniou2012,LI2010}, the classical first order Sobol index is split into various components. These components represent uncorrelative, interactive and correlative contributions of a given parameter to the output variance. However, the interpretation of these contributions, as well as of total order indices, remains unclear.
In this approach, the surrogate model is set up using independent joint input distribution. The polynomials of the PCE expansion are evaluated with the dependent samples, subsequently used to compute the covariance of the components functions. Analysis of covariance is then used to compute the resulting indices and their decomposition into the three components.

A new set of the indices for correlated inputs was introduced by Mara and Tarantola~\cite{MARA2021,MARA2012,MARA2015}. 
Two distinct indices represent correlated and uncorrelated contributions of a given variable. These allow to distinguish between the mutual dependent contribution and the independent contribution of the parameter to the model response variance.
The dependent parameters are decorrelated using the Gram--Schmidt procedure and Rosenblatt transformation, such that standard SA methods such as PCE or QMC frameworks can be used. However, since the SA is no longer performed using the original parameters, additional attention needs to be put to interpretation of the sensitivity indices. Additionally, different permutations of the decorrelated variables can be obtained, thus resulting in multiple set of the indices.

\subsection{Contribution and Organization}

SA with correlated parameters is studied in this work. The decorrelation approach is based on transformation of the input parameter space, such that	the SA is performed using the independent distributions, following the approach (ii) and the work of Mara and Tarantola~\cite{MARA2021,MARA2012,MARA2015}. 
The contributions are the following:
\begin{itemize}
    \item The correlated SA approach is studied in context of both variance-based and derivative-based sensitivities;
    \item Two transformations are used in order to reflect the stochastic dependencies in the input parameters, Cholesky decomposition of the correlation matrix and the Rosenblatt transformation;
    \item The methods are implemented within EasyVVUQ  SA framework, aiming to leverage large-scale computational resources to make state-of-the-art uncertainty quantification algorithms available and accessible to a wide range of computational scientists;
    \item Demonstrate the importance of the parameter correlations in the SA and provide extensive numerical experiments accompanied by a comprehensive interpretation of the results.
\end{itemize}

The following Sec.~\ref{sec:SA} discusses various aspects of the  SA, introducing both variance and derivative based indices. The treatment of the correlated variables and modifications of the SA algorithm are introduced in Sec.~\ref{sec:dependency}. The application model used in the numerical experiments is presented in Sec.~\ref{sec:models}. Finally, extensive numerical experiments and their analysis is provided in Sec.~\ref{sec:experiments}. The paper concludes in Sec.~\ref{sec:conclusions} outlining also future research directions.

	\section{SA Method without Correlations \label{sec:SA}}
The model is usually a complex interaction between its input parameters and outputs, and is treated in a black box fashion for the purpose of non-intrusive SA. Consider a model $U$ that is defined over a time horizon $\bft$, space dimension $\bfx$ and a set of $D$ uncertain  input parameters $\Q = \{Q_1, Q_2,\ldots\, Q_D\}$, such that
\begin{linenomath}
\begin{equation}
    \label{eq:model}
    \Y = U(\bft,\bfx,\Q).
\end{equation}
\end{linenomath}
The model includes uncertain parameters that can be collectively described by a joint multivariate probability density function $\rho_\Q$. If the uncertain parameters are statistically independent, the multivariate probability density function $\rho_\Q$ can be defined by separate univariate probability density functions $\rho_{Q_i}$, one for each uncertain parameter $Q_i$,
\begin{linenomath}
\begin{equation}
    \rho_\Q = \prod_{i=1}^{D} \rho_{Q_i}, \label{eq:marg_prod}
\end{equation}
\end{linenomath}
where unit normal distributions are assumed, such that $\rho_{Q_i} \sim \mathcal{N}(\mu=0, \sigma=1)$.

The main computational pattern of the SA in both MC and PCE consists of drawing the samples $\bfq$ from the input parameter space $\rho_\Q$ and evaluating the model $U(\bft,\bfx,\bfq)$ at these points. 
The number $N$ of such evaluations in the PCE approach
\begin{linenomath}
\begin{equation}
    N = \begin{pmatrix}
        D+P \\
        P
    \end{pmatrix}
\end{equation}
\end{linenomath}
is a function of the polynomial degree $P$ of the basis and the dimension $D$  of the parameters, where $N$ grows fast, especially with the increasing dimension of the parameters.
Based on these model evaluations, the true response of the model $\Y$ is fitted onto a polynomial basis $\Psi = \{\Psi_p, p=0,\ldots,P\}$ with a polynomial degree up to $P$. The basis needs to be orthogonal with respect to the input distributions $\rho_{Q_i}$. The polynomial model $\hat{\Y} = \hat{U}(\bft,\bfx,\Q)$ is build such that the true model is approximated by the polynomial expansion,
$U(\bft,\bfx,\Q) \approx \hat{U}(\bft,\bfx,\Q)$,
and the model outputs are similar $\Y \approx \hat{\Y}$. The surrogate model $\hat{U}(\bft,\bfx,\Q)$ is built from the polynomial basis $\Psi$ as
\begin{linenomath}
\begin{align}
 \hat{U}(\bft,\bfx,\Q) &= \sum_{p\subset P} a_p \Psi_p(\Q) \nonumber  \\
                & = a_0 \Psi_0 + \sum_{p\subset P} \sum_{i=1}^D a_p^i \Psi_p^i(Q_i) \nonumber  \\
                &+\sum_{p\subset P}  \sum_{i,j=1, j>i}^D  a_p^{ij} \Psi_p^{ij}(Q_i,Q_j)  \nonumber  \\
                & \hspace{1cm} \vdots \nonumber  \\
                &+  \sum_{p\subset P} a_p^{12\ldots D} \Psi_p^{12\ldots D}(Q_1,\ldots,Q_D),
                \label{eq:components}
\end{align}
\end{linenomath}
where $\Psi_0 = 1$ is a zero order polynomial, $\Psi_p^i(Q_i) $ is a single dimensional polynomial up to degree $p$ for a single input $Q_i$, $\Psi_p^{ij}(Q_i,Q_j) $ denotes polynomial  order up to $p$ of combination of two inputs $Q_i,Q_j$, etc. The polynomial coefficients $a_p$ follow similar notation.
In the non-intrusive variant of the method, the polynomial basis $\Psi$ is constructed using, e.g., the three terms recurrence or the discretized Stieltjes method~\cite{chaospy,EasyVVUQ}. The orthogonality of the polynomials holds in case the $\Q$ parameters are independent, i.e., the joint density can be expressed as a product of the individual marginal densities from Eq.~\eqref{eq:marg_prod}.

A set of the polynomial coefficients $a_p$ is determined such that the PCE model $\hat{U}$ approximates the true model response $\Y$. In point collocation, the approximation is built such that it minimizes the error at a set of collocation nodes compared to the true model response. Hammersley sampling~\cite{chaospy} from the distribution is used to choose the collocation points. This results in a set of linear equations for the polynomial coefficients, which are solved using e.g. Tikhonov regularization. The overall algorithm is summarized in Alg.~\ref{alg:SA}, where the SA is described in the following sections.

\begin{algorithm}[t!]
	\caption{SA Method without Correlations.}\label{alg:SA}
	\begin{enumerate}
		\item Generate samples $\bfq_1, \dots, \bfq_N$ from the independent multivariate distribution $\rho_\Q$.
		\item Evaluate the true model $\Y_1=U(\bfx,\bft, \bfq_1), \dots, \Y_N=U(\bfx,\bft, \bfq_N)$ at $\bfq_i \in \rho_\Q$.
		\item Create a polynomial expansion $\Psi_1, \dots, \Psi_P$ up to  the $P$-th degree  from $\rho_\Q$.
		\item Solve the linear regression problem: $\Y_n = \sum_p a_p\ \Psi_p(\Q_n)$ for $a_1, \dots, a_p$.
		\item Construct the model approximation $U(\bfx, \bft, \Q) \approx \hat{U}(\bfx, \bft, \Q) = \sum_p a_p\ \Psi_p(\Q)$
		\item Perform the SA using the surrogate model $\hat{U}(\bfx, \bft, \Q)$.
	\end{enumerate}
\end{algorithm}

\subsection{Variance-based Sensitivity \label{sec:indices-var}}
Variance-based SA~\cite{saltelli2004sensitivity} determines the impact of the input parameters which can be used to asses the role of the parameters in the model, i.e., determine if the parameter contributes intrinsically or via the parameter interactions, or asses the relative importance of the individual parameters. 
Additionally, variance-based sensitivity quantifies the output uncertainty and its propagation through the model from the uncertain inputs~\cite{uncertainpy,EasyVVUQ}.
Following the variance decomposition~\cite{SOBOL2001}, the total output variance $V(Y_n)$ of $n$-th model output from Eq.~\eqref{eq:model} can be decomposed as
\begin{linenomath}
\begin{equation}
    V(Y_n) = \sum_i V_i + \sum_i\sum_{j>i} V_{ij} + \ldots + V_{12\ldots D},
\end{equation}
\end{linenomath}
where the partial variances are defined as
\begin{linenomath}
\begin{align}
    V_i &= \mathbb{V}(\mathbb{E}(Y_n|Q_i)), \\
    V_{ij} &= \mathbb{V}(\mathbb{E}(Y_n|Q_i,Q_j)) -V_i - V_j,
\end{align}
\end{linenomath}
and so on, and the total variance is $V(Y_n) = \mathbb{V}(\mathbb{E}(Y_n))$. The polynomial coefficients can be post-processed to compute quantities of interest such as mean, variance and other statistical moments or variance-based sensitivity indices~\cite{SUDRET2008,Caniou2012}.
The sensitivity indices in the variance-based measures, known as Sobol indices~\cite{SOBOL2001}, are defined as the fraction of the variance of the component functions with respect to the total variance. The first order sensitivity index $S_i$ measures the contribution of the $i$-th parameter,
\begin{linenomath}
\begin{equation}
    S_i = \frac{V_i}{V(Y_n)}. \label{eq:sobol}
\end{equation}
\end{linenomath}
The total order sensitivity index $S_i^T$ includes not only the intrinsic contribution of the parameter itself as is the case for the first order index, but also interactions with other parameters are considered,
\begin{linenomath}
\begin{equation}
    S^T_i = \frac{\sum_{\alpha}V_a}{V(Y_n)},
\end{equation}
\end{linenomath}
where $\alpha$ is a set of all multi-indices which contain $i$. It necessarily holds that
$0 \leq S_i \leq S^T_i \leq 1$,
and in case the model is additive and there are no parameter interactions, i.e. the higher order terms are zero, then
\begin{linenomath}
\begin{equation}
    \sum_i S_i = 1.
    \label{eq:sobol_sum}
\end{equation}
\end{linenomath}

\subsection{Derivative-based Sensitivity \label{sec:indices-der}}
Derivative-based sensitivity indices express how much does the model output change if a small perturbation is applied to some of the inputs. The analytical derivatives of the complex models are not known, thus the usual practice is to use automatic differentiation tools or adopt approximations techniques such as finite differences to evaluate  the numerical derivatives. The model derivative with respect to the parameter $Q_i$ at a fixed point ${Q_i^0}$ is expressed as 
\begin{linenomath}
\begin{equation}
  S^\mathcal{D}_i =   \left. \frac{\partial Y_n}{\partial Q_i}\right|_{Q_i^0}.
  \label{eq:derivative}
\end{equation}
\end{linenomath}
The shortcoming of this approach is that the resulting index can be computed only in the vicinity of the operating point of the given model configuration or its applicability for the SA of a single variable at a time, ignoring any possible interactions between the parameters.

Alternatively, the derivative-based sensitivity index can be evaluated by constructing the surrogate model $\hat{U}$ and compute the derivative of the polynomial expression with respect to a given parameter. With this approach, the interaction of the parameters can be incorporated in the SA via the parameter correlations. Thus, the sensitivity indices of the individual parameters can incorporate interaction with other parameters using the  procedure proposed in this paper. This approach can be used for both variance-based and derivative-based sensitivities.


\begin{figure}[t!]
	\centering
	\includegraphics[width=0.49\textwidth]{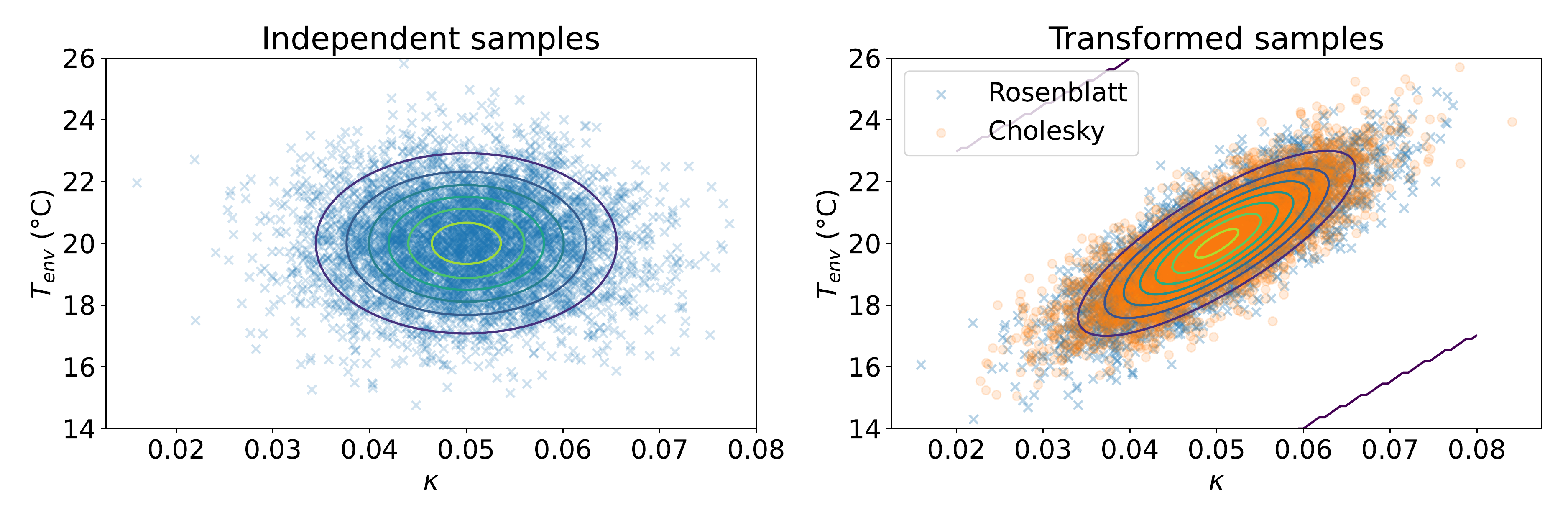}
	\caption{Independent normal distribution (left) of the input space and the corresponding transformed parameter space with $\rho_\mathcal{C}=0.8$ (right). The contour lines illustrate the multivariate probability density function.}
	\label{fig:rosenblatt}
\end{figure}

\section{SA Method with Correlations \label{sec:dependency}}

When considering models with correlated parameters,
the polynomial expansion~\eqref{eq:components} cannot be used to accurately represent the model sensitivity since it doesn't distinguish  whether the parameter is contributing to the model directly or through a correlation with another variable.
This can lead to incorrect conclusions about the variance-based decomposition, where the importance of the input parameters to the model and the sensitivity of the model to variations in these parameters no longer reflects the true parameter interactions in the model.

In order to address the parameter dependency, the parameters must be decorrelated prior to applying the SA. This approach is adopted in the procedure of Mara and Tarantola~\cite{MARA2021,MARA2012,MARA2015}.  In their original work, the samples are drawn from the correlated joint distribution and define a set of new variables, which are characterized by the conditional probability density functions and as such can be treated as independent.
In this work, the collocation points are sampled using the independent unit normal distributions $\rho_\Q = \N^D(\bfmu, I)$, while the model is evaluated using the transformed samples considering also the dependencies.  Fig.~\ref{fig:rosenblatt} illustrates this principle, the independent collocation nodes and their transformation to the target correlated distribution $\rho_\Q^* = \N^D(\bfmu, \C)$.  Since the linear relationship between the random variables is characterized using the Pearson and Spearman correlation coefficients, the correlated samples can be obtained from the independent ones using two different methods; (i) Rosenblatt transformation~\cite{Rosenblatt-original} 
 and (ii) Cholesky decomposition of the correlation matrix~\cite{choleskyTransform}.

\begin{algorithm*}[t!]
	\caption{SA Method with Correlations.}\label{alg:SAcorrelated}
	Generation of samples and their transformation:
	\begin{enumerate}
		\item Generate samples $\bfq_1, \dots, \bfq_N$ from the independent multivariate distribution $\rho_\Q$.
		\item Transform the samples $\bfq_i \in \rho_\Q $ to $\bfq^*_i \in \rho^*_\Q,\ i = 1, \dots, N$, using:
		\begin{enumerate}
			\item Cholesky transformation $\bfq^*_i = T(\bfq_i) = \bfq_i L$ from~\eqref{eq:cholesky_transform}.
			\item Rosenblatt transformation $\bfq_i^* = T(\bfq_i)$          from~\eqref{eq:rosenblatt_transform}.
		\end{enumerate}
	\end{enumerate}
	Construction of the surrogate model:
	\begin{enumerate}
		\item Evaluate the true model $\Y_1^*=U(\bfx,\bft, \bfq^*_1), \dots, \Y_N^*=U(\bfx,\bft, \bfq^*_N)$ at $\bfq^*_i \in \rho^*_\Q$.
		\item Create a polynomial expansion $\Psi_1, \dots, \Psi_P$ up to the $P$-th degree from $\rho_\Q$.
		\item Solve the linear regression problem: $\Y_n^* = \sum_p a_p\ \Psi_p(\Q_n)$ for $a_1, \dots, a_p$.
		\item Construct the model approximation $U(\bfx, \bft, \Q^*) \approx \hat{U}(\bfx, \bft, \Q) = \sum_p a_p\ \Psi_p(\Q)$.
	\end{enumerate}
\end{algorithm*}

\subsection{Cholesky Decomposition}

Independent samples with an identity correlation matrix are drawn from a joint multivariate distribution
\begin{linenomath}
\begin{equation}
    \Q \sim \N^D(\bfmu, I).
\end{equation}
\end{linenomath}
Since the components $Q_i$ are random variables with zero mean and unit variance with zero correlation, we have $\mathbb{E}(Q_iQ_j) = \delta_{ij}$. Hence, $\mathbb{E}(\Q\Q^{T}) = I$. The joint probability of the independent variables can be expressed as the product of the marginal distributions. On the other hand, the joint distribution of the dependent variables
\begin{linenomath}
\begin{equation}
    \Q^* \sim \N^D(\bfmu, \C).
\end{equation}
\end{linenomath}
 can be expressed as a product of the conditional distributions, which are not known. An alternative approach is it to introduce a transformation between the two spaces of the variables, such that the independent variables $\Q$ can be transformed to $\Q^*$ and vice versa. The transformation is defined via the Cholesky decomposition of the correlation matrix. The Cholesky decomposition of the correlation matrix $C$ is computed such that
$L = \text{chol}(C)$, and $LL^T = C$, where

\begin{linenomath}
\begin{equation}
    L = \begin{pmatrix}
        c_{11} & & \\
        c_{21} & c_{22} & \\
        c_{31} & c_{32} & c_{33}
    \end{pmatrix}.
\label{eq:cholesky_transform}
\end{equation}
\end{linenomath}

The uncorrelated samples $\Q$ are then transformed to samples that contain the correlations between the variables, as given by the correlation matrix, such that the transformed samples behave as drawn from the correlated distribution, i.e., $\Q^* = T(\Q) = L\Q$, where $T$ is the transformation operator,
\begin{linenomath}
\begin{equation}
    \label{eq:T}
    \begin{pmatrix}
        Q^*_{1}\\
        Q^*_{2}\\
        Q^*_{3}
    \end{pmatrix}
    = \begin{pmatrix}
        c_{11} & & \\
        c_{21} & c_{22} & \\
        c_{31} & c_{32} & c_{33}
    \end{pmatrix}
    \begin{pmatrix}
        Q_{1}\\
        Q_{2}\\
        Q_{3}
    \end{pmatrix}.
\end{equation}
\end{linenomath}

The random vector $\Q^*$ behaves such that
$\mathbb{E}(\Q^*\Q^{*T}) = \mathbb{E}((L\Q)(L\Q)^T) = \mathbb{E}(L\Q\Q^{T}L^T) = L\mathbb{E}(\Q\Q^{T})L^T = LIL^T = C $, since expectation is a linear operator.
Hence, the transformed random vector $\Q^*$ has the desired correlation matrix $C$ and $\Q^* \sim \N^D(\bfmu, \C)$. 
%
One of the requirements for the Cholesky decomposition is that the matrix is positive definite. In practice, the sample covariance matrix is always at least positive semi-definite~\cite{CholeskyPD}. In certain situations, the eigenvalues of a covariance matrix can be zero. This can happen when the set of parameters includes constant or perfectly correlated variables, or the sample size is too small. In this work, the covariance matrix is always assumed to be positive definite.

\begin{algorithm*}[t!]
	\caption{Evaluation of the Sensitivity Indices.}\label{alg:PCEanalysis}
	Variance-based analysis:
	\vspace{-0.3cm}
	\begin{enumerate}
		\item Compute the variance-based indices $S_i$ from the coefficients of $\hat{U}(\bfx, \bft, \Q)$ according to~\eqref{eq:sobol}~\cite{SUDRET2008,Caniou2012}.
	\end{enumerate}
	Derivative-based analysis:
	\vspace{-0.3cm}
	\begin{enumerate}
		\item Compute the partial derivatives of the polynomial model $\hat{U}(\bfx, \bft, \Q)$ with respect to $Q_i$ according to~\eqref{eq:derivative}.
		\item Evaluate the derivatives at the point of interest, e.g. the mean value of the parameters, in order to obtain the sensitivity indices $S_i^\mathcal{D}$.
	\end{enumerate}
\end{algorithm*}

\subsection{Rosenblatt Transformation}
The Rosenblatt transformation~\cite{Rosenblatt-original} allows for a vector of independent random variables $\Q$ generated from the distribution $\rho_\Q$ to be transformed to the target distribution  $\rho_\Q^*$ which contains correlations between the variables. The transformed samples $\Q^* = T(\Q)$ behave as if they were drawn from the target density $\rho_\Q^*$. 

The Rosenblatt transformation can be derived  from a probability decomposition of a bivariate random variable $\boldsymbol{Q}^*=\left(Q_{1}^*, Q_{2}^*\right)$ with a correlation as
\begin{linenomath}
\begin{equation}
\rho_\Q^* = \rho_{Q_1} \rho_{Q_{2} \mid Q_{1}},
\end{equation}
\end{linenomath}
where $\rho_{Q_1}$ is a marginal density function, and $\rho_{Q_{2} \mid Q_{1}}$ is a conditional density. In a general multivariate case, the density decomposition has the form
\begin{linenomath}
\begin{equation}
\rho_\Q^* = \rho_{Q_1}  \prod_{{d_i}=2}^{D} \rho^{\prime}_{Q_{d_i}},
\end{equation}
\end{linenomath}
where
$\rho^{\prime}_{Q_{d_i}} = \rho_{Q_{d_i}} |\rho_{Q_1}, \ldots, \rho_{Q_{{d_{i-1}}}}$
is conditioned on all components with lower indices. A forward Rosenblatt transformation is then defined as
\begin{linenomath}
\begin{equation}
T = \left(F_{Q_{1}^{\prime}}, \ldots, F_{Q_{d}^{\prime}}\right),
\label{eq:rosenblatt_transform}
\end{equation}
\end{linenomath}
where $F_{Q_{{d_i}}^{\prime}}$ is the cumulative distribution function
\begin{linenomath}
\begin{equation}
F_{Q_{{d_i}}^{\prime}} = \int_{-\infty}^{q_{{d_i}}} \rho_{Q_{{d_i}}^{\prime}}\left(r \mid q_{1}, \ldots, q_{d_i-1}\right) \mathrm{d} r.
\end{equation}
\end{linenomath}
Note also that the Rosenblatt transformation is not limited to only Gaussian distributions. In this work, the implementation of the transformation implemented in the Chaospy~\cite{chaospy} package is used.


\subsection{SA Method with Correlations}

The SA algorithm introduced in Alg.~\ref{alg:SA} needs to be modified in presence of the correlated inputs in order to correctly represent the input-output interactions and the sensitivity indices. The changes are summarized in the modified method presented in Alg.~\ref{alg:SAcorrelated}.

The modified method first needs to generate the parameter samples including the correlations. As before, the set of the parameter samples $\bfq$ is generated from the independent joint distribution $\rho_\Q$ which are subsequently transformed according to the stochastic dependency structure. The correlated samples $\bfq^* = T(\bfq) $ can be obtained using either the Cholesky or Rosenblatt transformations.

Having created the correlated samples, the modified method next evaluates the true model using the correlated samples $\Y^*=U(\bfx,\bft, \bfq^*)$. The surrogate model is constructed in the transformed coordinate space compared to the independent model, reflecting the correlated contributions which affect the model outputs. This coordinate space is transformed implicitly, by mapping the polynomial expansion generated from the independent distribution $\rho_\Q$, to the space of the correlated model outputs $\Y_n^*$. In other words, the linear regression 
\begin{linenomath}
\begin{equation}
	\Y_n^* = \sum_p a_p(t)\ \Psi_p(\bfq_n)
\end{equation}
\end{linenomath}
is solved, where the left-hand side term is in the correlated space, while the polynomial expansion and the samples $\bfq_n$ at the right-hand side is from the uncorrelated space. Such surrogate model is then used to perform the SA summarized in Alg.~\ref{alg:PCEanalysis}. 

 \begin{table*}[t!]
	\caption{Sensitivity indices for different parameter permutations $\mathcal{P}_i$.}
	\label{tab:indices}
	\centering
	\begin{tabular}{l|c|l|c}
		Permutation & Full Index \hskip1em& Marginal Indices \hskip1em & Independent Index \\ \cline{1-4}
		$\mathcal{P}_1 = (1,2,3,\ldots, D)$              &     $Q_1$       &   $Q_2,\ldots, Q_{D-1}$            &     $Q_D$       \\
		$\mathcal{P}_2 = (2,3,\ldots, D, 1)$          &     $Q_2$       &   $Q_3,\ldots, Q_{D}$               &     $Q_1$        \\
		$\mathcal{P}_3 = (3,\ldots, D, 1,2)$       &     $Q_3$       &   $Q_4,\ldots, Q_{D}, Q_1$       &     $Q_2$        \\
		\hskip1.6em \vdots               &       \vdots      &   \hskip1.6em \vdots                                     &     \vdots        \\
		$\mathcal{P}_D = (D, 1,2,\ldots,D-1)$    &     $Q_D$      &   $Q_1,\ldots, Q_{D-2}$            &     $Q_{D-1}$          
	\end{tabular}
\end{table*}

\subsection{Interpretation of the Sensitivity Indices \label{sec:permutations}}
The sensitivity indices computed following the method presented in Alg.~\ref{alg:SAcorrelated}~and~\ref{alg:PCEanalysis} need to be interpreted differently, compared to their counterparts computed without any parameter dependencies (see Sec.~\ref{sec:indices-var} and~\ref{sec:indices-der}). One needs to consider the fact that the parameter transformations effectively introduce new variables, which are a combination of the original ones in case of linear dependencies. Consequently, the resulting indices either include the effects of the parameter itself together with its dependence with other inputs or it can represents the sensitivity index without its mutual dependent contributions with other parameters.

When applying the transformation a particular ordering of the parameters is assumed, e.g., the natural ordering $\mathcal{P}_1 = (1,2,\ldots, D)$ with the parameters $\mathcal{P}_1 \Q = (Q_1, Q_2,\ldots\, Q_D)$. The transformation is then applied sequentially, where the first parameter is kept unmodified, while the others are transformed according to the particular correlation structure. Considering a vector of the input parameters $\mathcal{P}_1 \Q$, the correlated vector is formed as
\begin{linenomath}
\begin{align}
    Q_1^* &= Q_1, \nonumber \\
    Q_2^* &= Q_2|Q_1,  \nonumber\\
    Q_3^* &= Q_3|Q_1Q_2,  \\
                & \vdots \nonumber \\
    Q_D^* &= Q_D|Q_1Q_2\ldots Q_{D-1}. \nonumber
\end{align}
\end{linenomath}
The resulting sensitivity indices obtained by applying the SA with correlations using the transformed samples $\Q^* = (Q_1^*, Q_2^*,\ldots\, Q_D^*)$ need to be interpreted differently, since different variables have been used compared to the original variables $\Q$. 
One needs to distinguish between the Full and Independent indices. The Full index includes the effects of the parameter itself together with its dependence with all other inputs. On the other hand, the Independent index represents the contribution of a parameter without its mutual dependent interactions with other parameters. Using the permutation $\mathcal{P}_1$, the Full index for the parameter $Q_1$ is obtained, together with the Independent index for the parameter $Q_D$.
The Full index is obtained for the first parameter in the permuted vector $\mathcal{P}_1 \Q$, while the independent index corresponds to the last parameter in the permuted vector.  The sensitivity indices of the remaining variables in the vector $\mathcal{P}_1 \Q$, that is $( Q_2,\ldots\, Q_{D-1})$, express the marginal contribution of $Q_i,\ i=2,\ldots,D-1$ to the output variance without its correlative contributions with parameters $Q_j,\forall j: j < i$.
Thus, under the permutation $\mathcal{P}_1$ the Full index for the parameter $Q_1$ is defined as 
\begin{linenomath}
	\begin{equation}
		S_1 = \frac{\mathbb{V}(\mathbb{E}(Y_n|Q^*_1))}{\mathbb{V}(Y_n)},
	\end{equation}
\end{linenomath}
while the Independent index for the parameter $Q_D$ is defined as 
\begin{linenomath}
	\begin{equation}
		S_D = \frac{\mathbb{V}(\mathbb{E}(Y_n|Q^*_D))}{\mathbb{V}(Y_n)}.
	\end{equation}
\end{linenomath}
Note that the Full index is computed for the parameter $Q^*_1=Q_1$ which is chosen from its marginal distribution $\rho_{Q_1}$ and that it carries mutual contributions to the total variance due to the dependence on other parameters $Q_j,\ j>1$. On the other hand, the Independent index for the parameter $Q^*_D=Q_D|Q_1Q_2\ldots Q_{D-1}$ does not contain the mutual contributions with other parameters, since the parameter was drawn from the conditional distribution $\rho_{Q_D|Q_{1}Q_{2}\ldots Q_{D-1}}$.

 In order to compute the remaining Full and Independent indices, different permutations need to be used. For example $\mathcal{P}_2 = (2,\ldots, D, 1)$, such that $\mathcal{P}_2\Q = (Q_2, Q_3,\ldots\, Q_D, Q_1)$ from which the Full index of parameter $Q_2$ and Independent index of $Q_1$ can be determined. Overall, there exist $D!$ different permutations. However, both indices for all parameters can be obtained by circularly reordering the input vector $\Q$, i.e., performing the SA $D$ times in total, as summarized in Tab.~\ref{tab:indices}.

	\section{Application Model  \label{sec:models}}
The coffee cup model~\cite{uncertainpy} simulates a cooling process of a liquid contained in an open container. The model uses Newton's law of cooling to evolve the temperature $T$ over the simulation time $t$,
\begin{linenomath}
\begin{equation}
    \frac{dT(t)}{dt} = -\kappa (T(t) -T_{env}).
    \label{eq:coffee_model}
\end{equation}
\end{linenomath}
The parameter $\kappa$ characterizes the container containing the liquid and the rate at which it dissipates the heat to the environment. Ambient temperature of the environment is represented by the parameter $T_{env}$, while the initial temperature of the liquid is specified by the constant $T_0 = 95 ^\circ\text{C}$.

In this study, the SA of the $\kappa$ and $T_{env}$ parameters is studied. Due to the measurement error, insufficient knowledge of the physical model or other reasons, the parameters $\kappa$ and $T_{env}$ cannot be assigned an exact numerical value representing the modeled physical system. Instead the parameters are modeled as uncertain and they are described with probability distributions. A normal distribution $\mathcal{N}(\mu,\sigma)$ is assumed in this work, with a given mean $\mu$ and standard deviation $\sigma$ for each parameter,
\begin{linenomath}
\begin{equation}
	\begin{aligned}
		\kappa &=  \mathcal{N}(0.05, 0.008), \\
		T_{env} &=  \mathcal{N}(20, 1.5).
	\end{aligned}
	\label{eq:coffee_model_params}
\end{equation}
\end{linenomath}
 
On top of the uncertainty in the individual parameters, these parameters might be correlated with each other. The correlation captures a physical property of the container's material and its heat transfer rate, witch changes depending on the ambient temperature of the environment. For example, as the ambient temperature $T_{env}$ increases, the material dissipates the heat more efficiently, increasing also the value of the parameter $\kappa$. The stochastic dependency of the two parameters is described using a correlation matrix $C$ with correlation between the parameters specified by $\rho_{\mathcal{C}}$ ,
\begin{linenomath}
\begin{equation}
	C = \begin{pmatrix}
	1.0 & \rho_{\mathcal{C}} \\
	\rho_{\mathcal{C}} & 1.0
	\end{pmatrix}.
	\label{eq:coffee_model_corr}
\end{equation}
\end{linenomath}
Fig.~\ref{fig:rosenblatt} illustrates the probability density function of the parameters, both with and without the correlation. The goal of the SA is to analyze the impact of the uncertain parameters to the outcome of the model, considering also the correlation between the parameters. 

%
%
	\begin{figure*}[t!]
	\centering
	\includegraphics[width=0.7\textwidth]{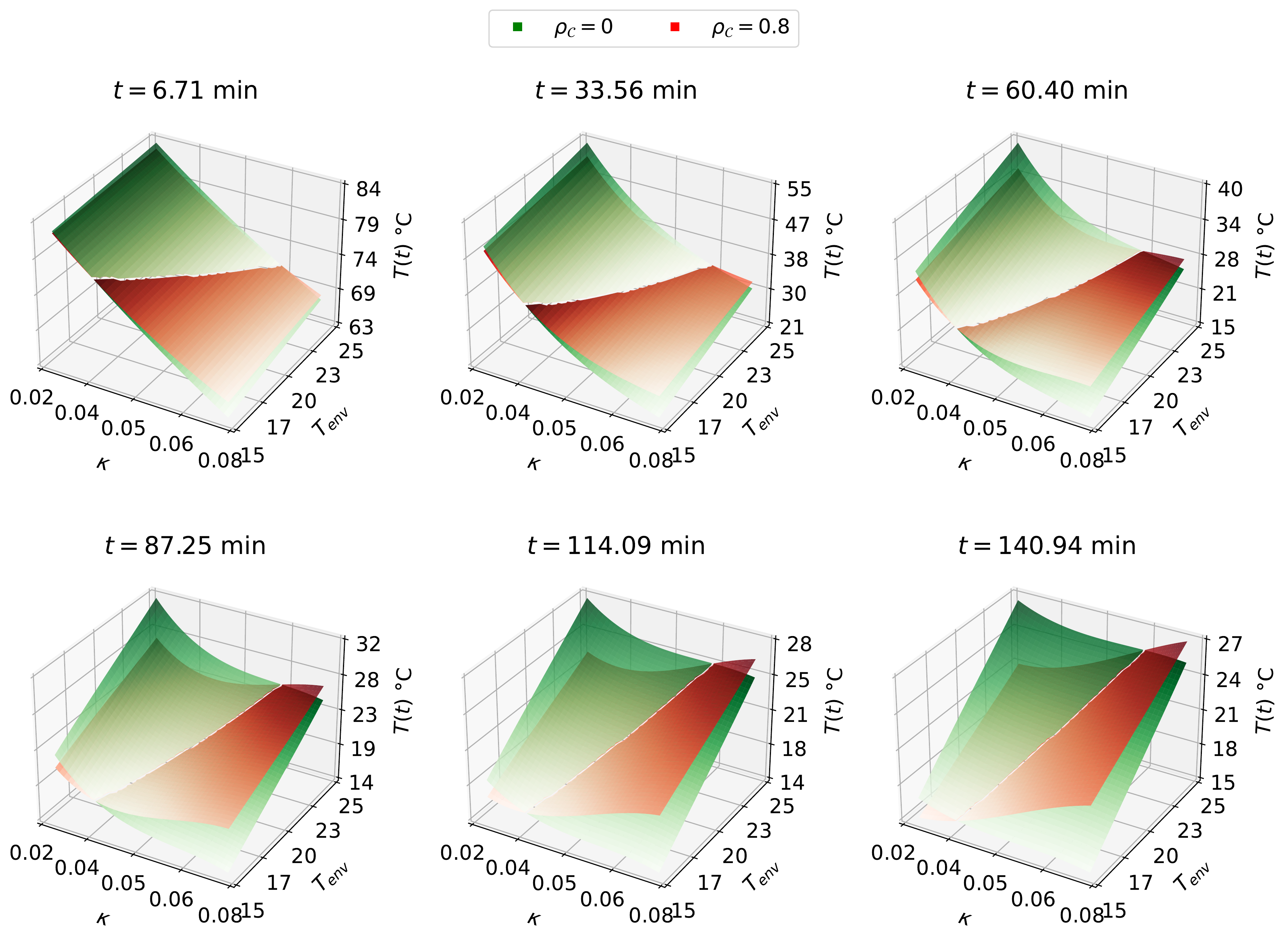}
	\caption{Surrogate models $\hat{U}_{\rho_{0}}$, $\hat{U}_{\rho_{\mathcal{C}}}$ for the coffee cup model  with independent and correlated inputs at various time instants.
		\label{fig:coffee_models}}
\end{figure*}

\begin{figure*}[t!]
	\centering
	\includegraphics[width=0.8\textwidth]{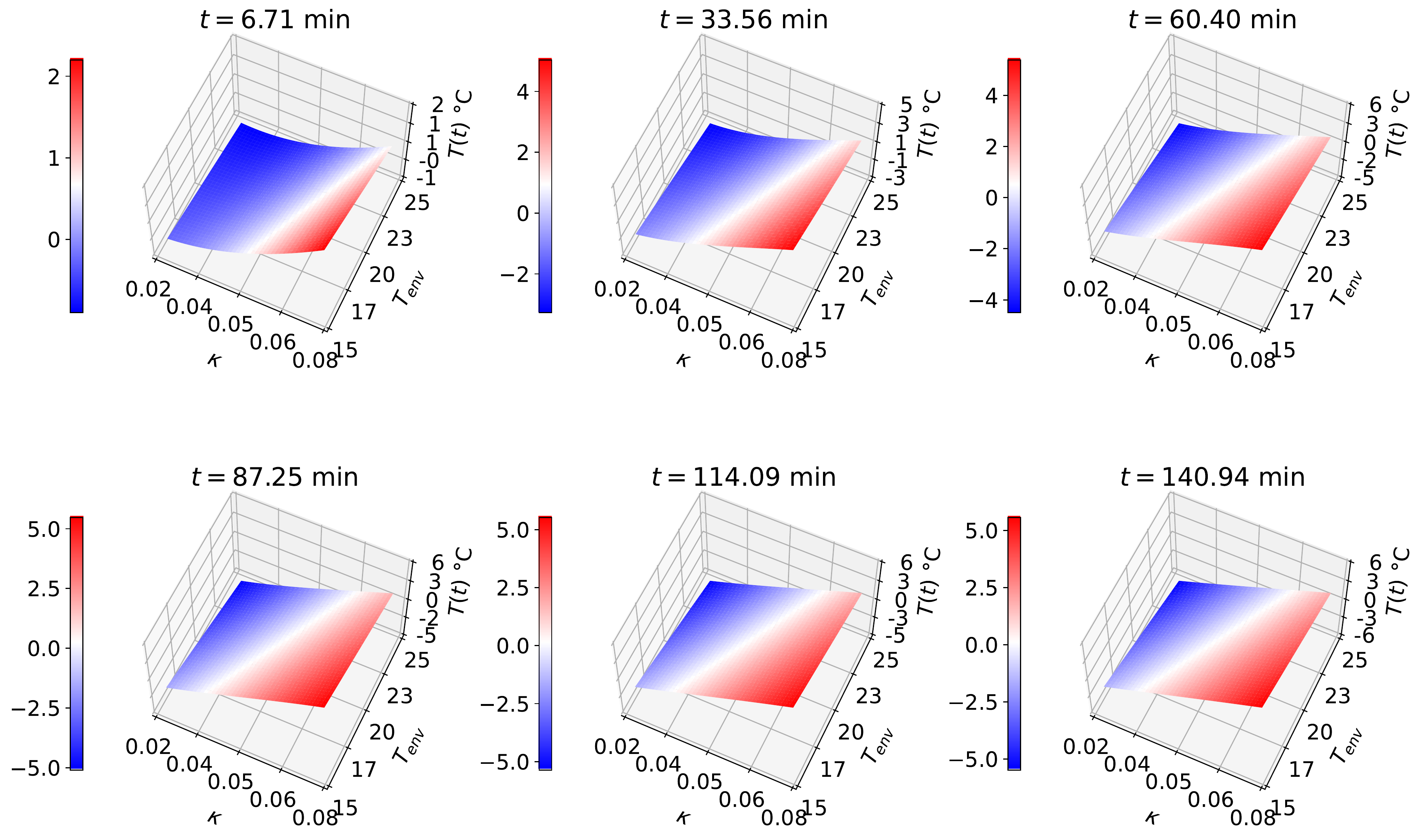}
	\caption{Difference between the surrogate models $\hat{U}_{\rho_{0}}$ and $\hat{U}_{\rho_{\mathcal{C}}}$ with independent and correlated inputs at various time instants.
		\label{fig:coffee_models_err}}
\end{figure*}

\subsection{Software Tools and Libraries \label{sec:swtools}}
The VECMA toolkit, or VECMAtk \cite{VECMA}, is used to manage the simulations required for the analysis. It enables automated verification, validation and UQ for complex applications, irrespective of their source domain. VECMAtk is optimized for large scale computations, and can be deployed on emerging high-performance computing (HPC) platforms. The toolkit has previously been used for a range of applications, such as a COVID model~\cite{covid} (with computational complexity in order of $10^4$ core hours per experiment), a molecular dynamics model~\cite{molecular} (experiments consumed $2\cdot 10^6$ core hours), and a range of other applications~\cite{VECMAapps}.


The EasyVVUQ package~\cite{EasyVVUQ}, a component of the VECMA toolkit, has been developed to facilitate forward UQ for HPC applications. EasyVVUQ supports the definition of custom UQ and SA procedures, which may include sampling and analysis, without requiring users to modify their core applications. It has been applied successfully to a diverse set of applications, and is able to cope with procedures that require thousands of simulation runs. EasyVVUQ is open source and written in Python~3.

	\section{Numerical Experiments \label{sec:experiments}}
Numerical experiments are performed using the model introduced in Sec.~\ref{sec:models}. 
The initial condition for the differential equation~\eqref{eq:coffee_model} used hereafter is $T_0 = 95 ^\circ\text{C}$.  The simulation time covers first $t=200$ minutes of the cooling process, with the time discretized into $150$ time steps of length $\Delta t = 80\,s$.
The parameter distributions used in the  numerical experiments, if not stated otherwise, are defined in Eq.~\eqref{eq:coffee_model_params}~and~\eqref{eq:coffee_model_corr}. The surrogate model is constructed using polynomials up to the third order, unless specified otherwise.

\subsection{Surrogate Models \label{sec:surrogates}}

The polynomial surrogates of the model~\eqref{eq:coffee_model} are examined in the vicinity of the mean value of the parameters~\eqref{eq:coffee_model_params}. The surrogate model is built for each time instant of the discretized time horizon, depicting the model output as a function of the particular values of the  input parameters.
The surrogate models at various time instants $t$ for the coffee cup are illustrated in Fig.~\ref{fig:coffee_models}, demonstrating the effect of the correlation in the parameters. Note that while the difference between the two models is small near the begging of the simulation time, the gap between the two grows as the time progresses. The changes in the final temperature profile are exaggerated by the interaction of the parameters within the model over time, thus the effect of the correlation is particularly visible at advanced simulation time, i.e., $t>20-30\, min$.
The absolute difference between the uncorrelated and correlated surrogate models $e = \hat{U}_{\rho_{\mathcal{C}}} - \hat{U}_{\rho_{0}} $ with $\rho_{\mathcal{C}}  = 0.8$ is illustrated in Fig.~\ref{fig:coffee_models_err}.

It is also important to highlight different curvature of the surrogate models, since during the derivative-based analysis a partial derivative of the surrogate with respect to a parameter is evaluated at the mean value of the parameters. Similarly as before, the curvature difference between the two models 
$\hat{U}_{\rho_{\mathcal{C}}}$, $\hat{U}_{\rho_{0}}$ grows with the proceeding simulation time. 

\subsection{SA with Uncorrelated Parameters \label{sec:uncorrelated}}
\begin{figure}[t!]
    \centering
    \includegraphics[width=0.49\textwidth]{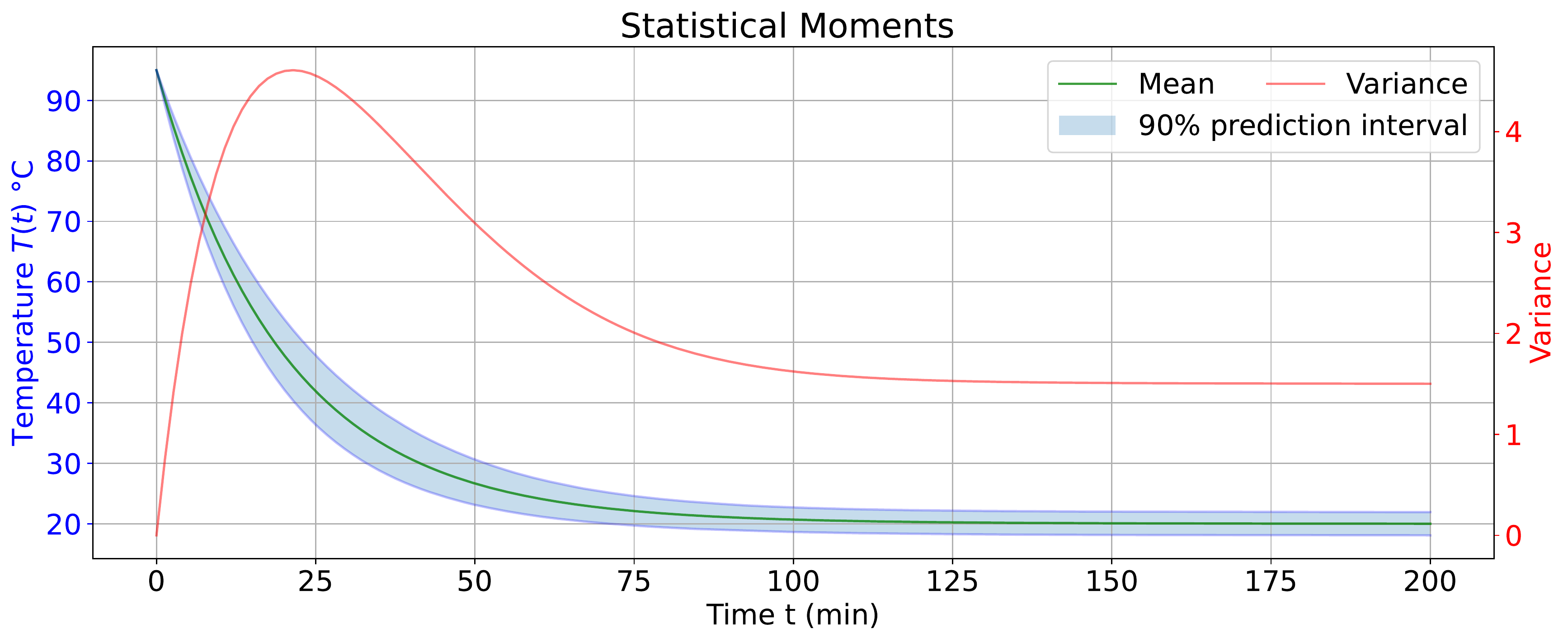}
    \caption{Statistical moments of the coffee cup model.
    \label{fig:coffee_independent_model}}
\end{figure}

In case the correlation matrix $\mathcal{C}$ is an identity matrix, i.e., there is no correlation between the parameters, the model evolution is shown in Fig.~\ref{fig:coffee_independent_model}. The model variance due to the uncertainty in the input parameters is shown as well. Note that the model variance at the initial point is zero, thus the Sobol indices are not defined at this time instant.

\subsubsection*{Variance-based Indices}
\begin{figure}[t!]
	\centering
	\includegraphics[width=0.48\textwidth]{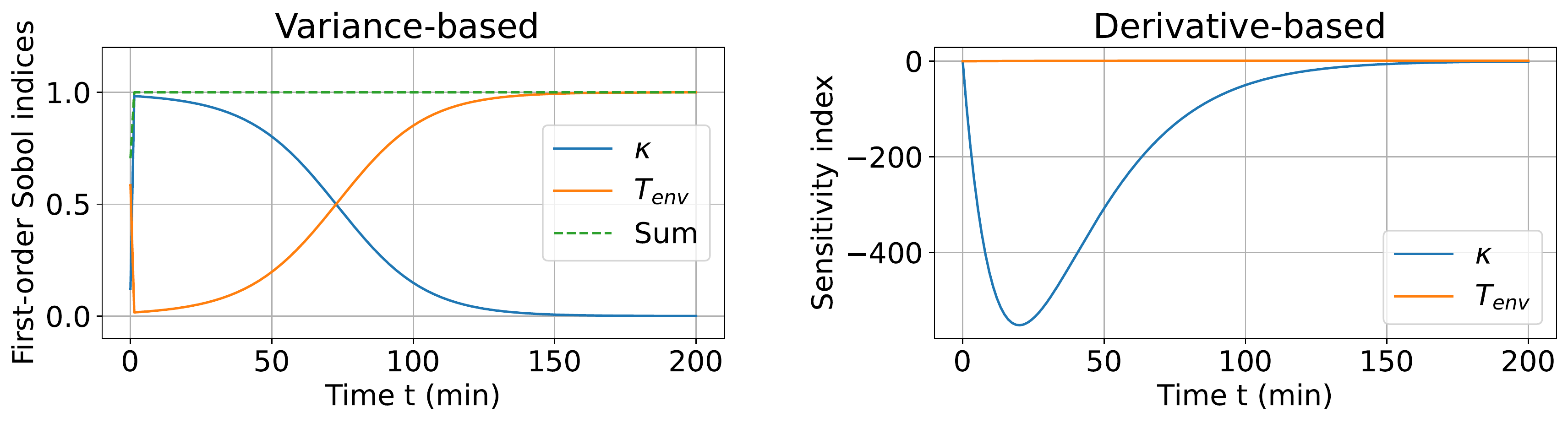}
	\caption{First-order Sobol and derivative-based indices considering the independent parameters.}
	\label{fig:coffee_independent}
\end{figure}

The corresponding sensitivity indices are shown in Fig.~\ref{fig:coffee_independent}, replicating the values of the variance-based Sobol indices from previous works, e.g.,~\cite{uncertainpy}. The difference of the first and total order variance-based indices, shown in the left panel of  Fig.~\ref{fig:coffee_independent}, are less than $10^{-4}$, indicating there are no higher order parameter interactions. The first order Sobol index of the $\kappa$ parameter is the most influential in the first 75 minutes, while the ambient temperature parameter dominates in the remaining simulation time. After reaching near equilibrium, i.e. the ambient and the coffee cup temperature difference is less than $\approx  0.1^\circ\text{C}$, the ambient temperature parameter explains nearly all of the output variance as shown in Fig.~\ref{fig:coffee_independent_model}. Intuitively, this is an expected behavior or the model, since the end state of the coffee cup after reaching the equilibrium is the environment temperature. 
Since there are no higher order interactions, the first order Sobol indices add up to one. Consequently, the behavior of the indices is necessarily complementary for the two parameters, i.e., if one index is increasing, the other is proportionally decreasing and vice versa.

\begin{figure}[t!]
	\centering
	\includegraphics[width=0.48\textwidth]{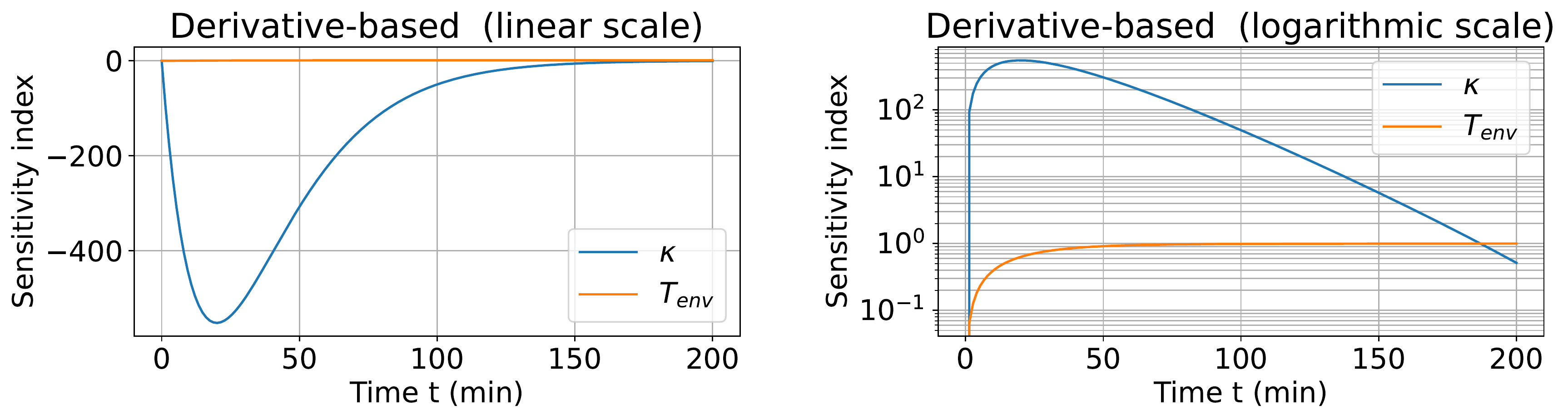}
	\caption{Derivative-based indices using linear and logarithmic y-axis scale. Note that the absolute values of the sensitivity indices are used in the latter case.}
	\label{fig:coffee_independent_log}
\end{figure}

\subsubsection*{Derivative-based Indices}
The derivative-based indices, shown in the right panel of  Fig.~\ref{fig:coffee_independent}, provide an insight into the model around the vicinity of a fixed point, in this case the mean value of the model parameters.  Following the definition in Eq.~\eqref{eq:derivative}, the values of the derivative-based index correspond to the slope of a tangent line to the model surface at the given spatial point and time instant. 
The magnitude of the individual derivative-based sensitivity indices differs by more than two orders of magnitude, thus the sensitivity indices are shown also in the logarithmic scale (considering their absolute values) in Fig.~\ref{fig:coffee_independent_log}. Note that the sensitivity of the parameter $T_{env}$ is near zero at the beginning of the simulation, which reflects the fact that it has very small contribution to the model output as the temperature of the coffee cup is driven mainly by the heat transfer constant. As the time progresses, the sensitivity of $T_{env}$ increases and approaches one, meaning that a change of the ambient temperature will have the proportional effect on the model output. This reflects the fact that the final temperature of the coffee cup is equal to the ambient temperature, thus the change in the ambient temperature induces an equal change in the final state of the coffee cup. On the other hand, the sensitivity of the parameter $\kappa$ is significantly larger but the sensitivity of the parameter decreases over the simulation time, since the heat transfer is driven mainly by the temperature gradient between the coffee cup and the surrounding environment which is largest in the begging of the simulation. As this temperature differential decreases, the heat transfer becomes less significant. Note also the negative value of the sensitivity index, meaning that as the heat transfer parameter $\kappa$ increases, the output of the model, that is the coffee cup temperature, decreases due to a larger effect of the heat transfer.


\subsection{SA with Parameter Dependency \label{sec:parameter_dep}}
Next, the correlation matrix $\mathcal{C}$ is modified, such that the off-diagonal elements are no longer zero, indicating parameter correlation. If not stated otherwise, the numerical experiments use the value $\rho_{\mathcal{C}} = 0.4$ for the Pearson correlation coefficient.  The ordering of the indices in SA with correlations becomes important and the SA needs to be performed for different permutations, as detailed in Sec.~\ref{sec:permutations}.
%

\begin{figure}[t!]
	\centering
	\begin{subfigure}[b]{0.485\textwidth}
		\centering
		\includegraphics[width=\textwidth]{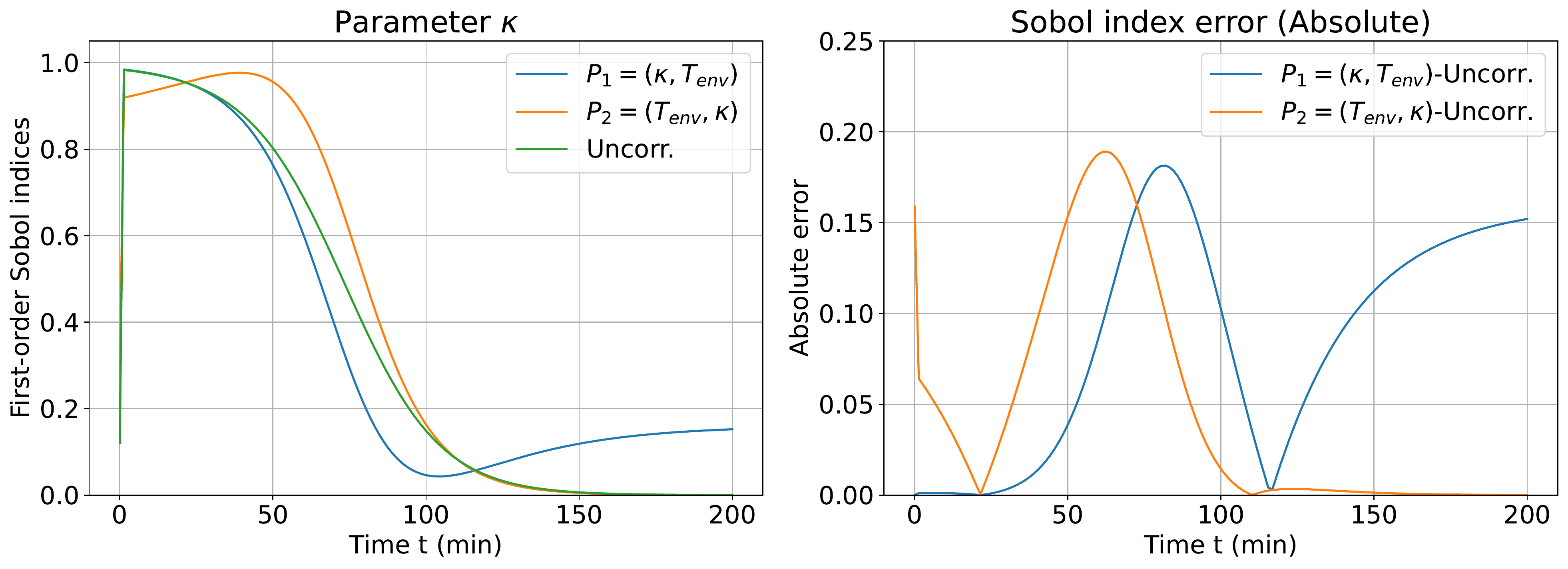}
		\caption{Full ($\mathcal{P}_1$) and Independent ($\mathcal{P}_2$) sensitivity indices of $\kappa$.}
		\label{fig:coffee_dependent_kappa}
	\end{subfigure}
	
	\begin{subfigure}[b]{0.485\textwidth}
		\centering
		\includegraphics[width=\textwidth]{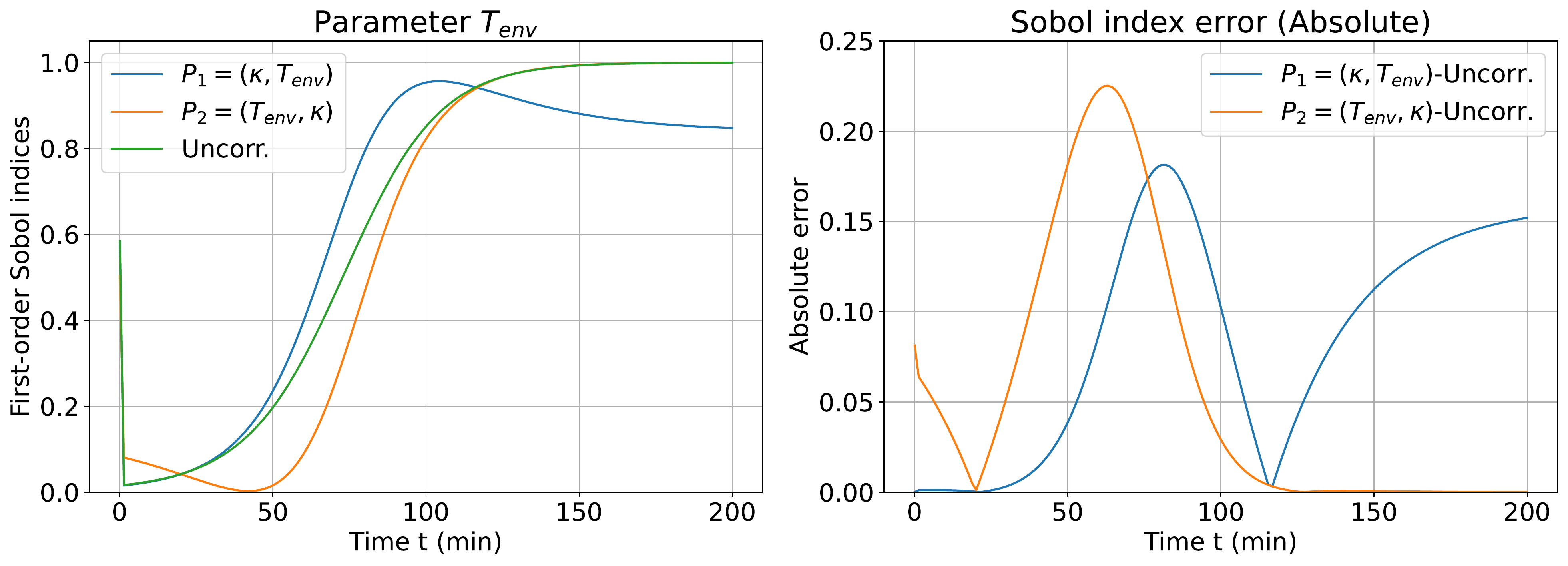}
		\caption{Full ($\mathcal{P}_2$) and Independent ($\mathcal{P}_1$) sensitivity indices of $T_{env}$.}
		\label{fig:coffee_dependent_tenv}
	\end{subfigure}
	\caption{Variance-based indices considering correlated parameters ($\rho_{\mathcal{C}}=0.4$) and the absolute difference with the uncorrelated indices from Fig.~\ref{fig:coffee_independent}.}
	\label{fig:coffee_dependent}
\end{figure}

\subsubsection*{Variance-based Indices}

The Sobol indices for the correlated parameters and their difference relative to the baseline experiment with independent parameters is shown in Fig.~\ref{fig:coffee_dependent}. 
In order to obtain the complete set of the  Full order and Independent indices for both parameters, the SA needs to be executed twice, each time with different parameter permutation. First the permutation $\mathcal{P}_1 = (\kappa, T_{env})$ is used to obtain 
the  Full order index of the $\kappa$ and the Independent index for the parameter $T_{env}$. Using the second permutation, $\mathcal{P}_2 = ( T_{env}, \kappa)$, the Full order index of the $T_{env}$ and the Independent index for the parameter $\kappa$ are obtained.

The Full order and Independent indices for the parameter $\kappa$ are shown in Fig.~\ref{fig:coffee_dependent_kappa}, comparing them to the sensitivity index shown in the previous section with uncorrelated parameters. 
Considering the  Full order index of the $\kappa$ parameter (permutation $\mathcal{P}_1$), the contribution of this parameter to the output variance near the end of the simulation time is increased compared to the independent case. Since the parameters are positively correlated, increasing the value of parameter $\kappa$ induces growth also in the $T_{env}$ parameter, thus increasing the end state equilibrium temperature of the coffee cup. Previously, there was no such interaction of the parameters, thus the variance-based index of the $\kappa$ parameter was zero at the end of the simulation time. 
However, the Full index is lower around the simulation time $t\approx 100\, min$ compared to the uncorrelated case. This is due to to the fact that increasing  $\kappa$ induces growth of the $T_{env}$ parameter, which in turn decreases the temperature gradient. Considering the dynamics of the model in~\eqref{eq:coffee_model}, the induced  of the ambient temperature counteracts the elevated heat transfer, thus the sensitivity of the heat transfer parameter has decreased.
When the Independent index is considered  (permutation $\mathcal{P}_2$), the effect of the correlation is removed and the index is nearly identical to the independent case in the simulation time $t > 100$min. However, in the simulation time around $t\approx 50$ the sensitivity of the $\kappa$ parameters has increased, thus emphasizing the importance of the parameter at the time instants when the temperature gradient is large.
Fig.~\ref{fig:coffee_dependent_kappa} also shows the absolute difference of the  Full order and Independent index compared to the uncorrelated case, in order to illustrate the magnitude of the difference.

The behavior of the first order Sobol index for $T_{env}$ parameter, as shown in Fig.~\ref{fig:coffee_dependent_tenv}, is opposite to that of the $\kappa$ parameter. The  Full order index (permutation $\mathcal{P}_2$) matches the independent case at the end of the simulation time (since it was already at the maximum value of 1). Removal of the contribution of the correlation decreases the value of the index. The magnitude of the change is proportional to the difference in the $\kappa$ parameter indices (Full vs. Independent index). 
%
This behavior of the indices can be interpreted such that the portion of the output variance can be explained by both parameters simultaneously since they are correlated. It can be equally said that some output variance is explained either by one or the other parameter. In an extreme case of the perfect correlation, $\rho_{12} = 1$, it is equivalent to say that the output variance is explained either by one or the other parameter, since the value of one parameter completely determines the value of the other.

\begin{figure}[t!]
	\centering
	\includegraphics[width=0.5\textwidth]{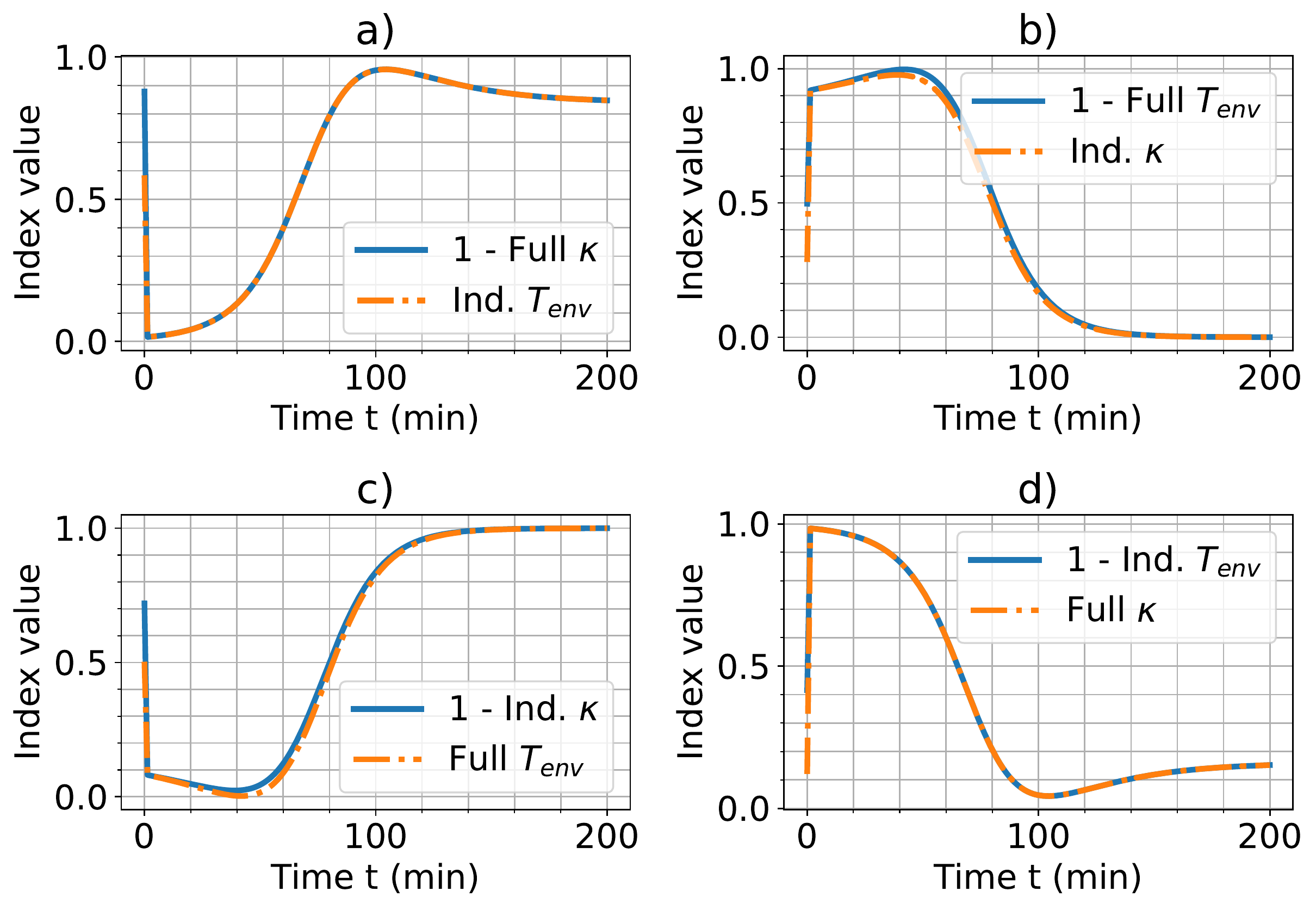}
	\caption{Complementary behaviors of the indices.}
	\label{fig:coffee_dependent_complement}
\end{figure}

It is also interesting to observe the complement of the indices to one, shown in  Fig.~\ref{fig:coffee_dependent_complement}. Consider the  Full index of the $\kappa$ parameter, as shown in Fig.~\ref{fig:coffee_dependent_complement}a. Its complement to one explains the output variance contributed by the other parameter alone without its correlated contribution with $\kappa$. In case of two parameters, this complement is the Independent Sobol index of the $T_{env}$ parameter. In general case with a set of $D$ parameters, complement of the  Full index of the parameter $i$ explains the amount of variance contributed by the remaining $D - 1$ parameters without their correlated contribution with~$i$.
A similar relationship is observed between the complement of the  Full index of $T_{env}$ and Independent index of $\kappa$ in Fig.~\ref{fig:coffee_dependent_complement}b. Note the presence of a numerical error in this case, in order to eliminate it, the indices should be computed with higher order polynomials in the PCE analysis (see Sec.~\ref{sec:convergence}).
%
Similarly, Fig.~\ref{fig:coffee_dependent_complement}c,d illustrate the relationship between the complement of the Independent indices and the  Full indices of the other parameter.

\subsubsection*{Derivative-based Indices}

\begin{figure}[t!]
	\centering
	\begin{subfigure}[b]{0.48\textwidth}
		\centering
		\includegraphics[width=\textwidth]{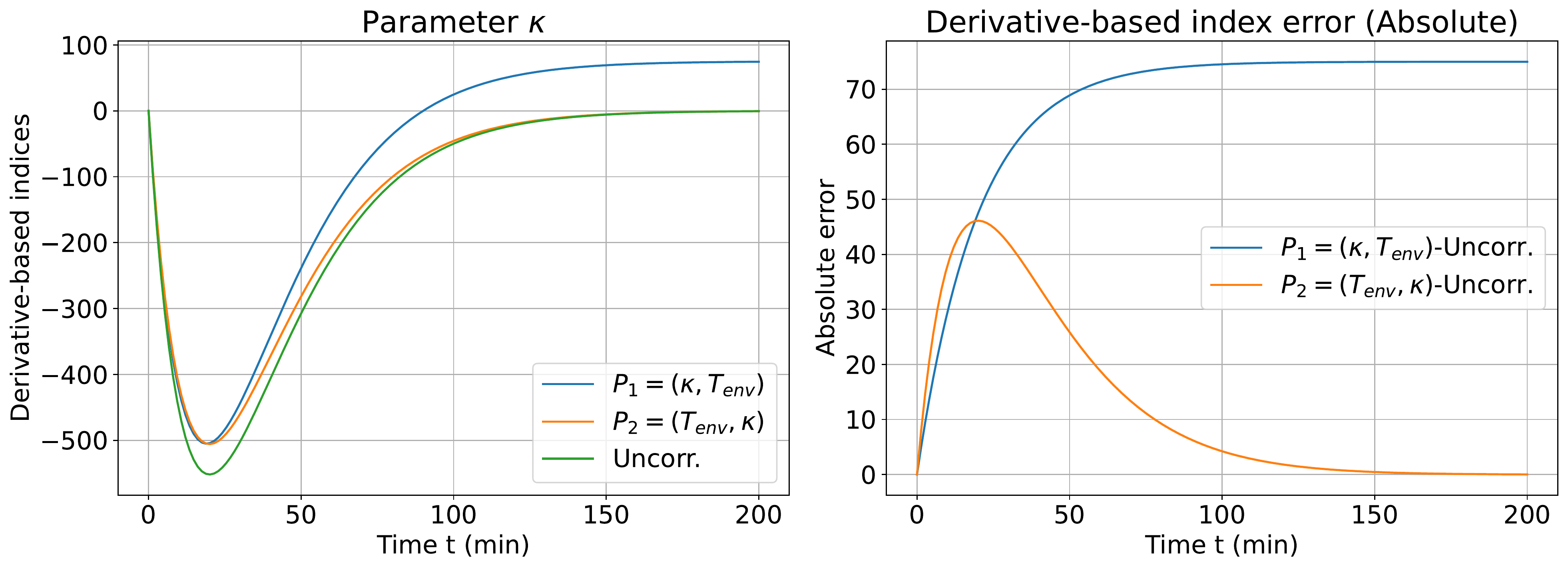}
				\caption{Full ($\mathcal{P}_1$) and Independent ($\mathcal{P}_2$) sensitivity indices of $\kappa$.}
		\label{fig:coffee_dependent_kappa_der}
	\end{subfigure}
	
	\begin{subfigure}[b]{0.48\textwidth}
		\centering
		\includegraphics[width=\textwidth]{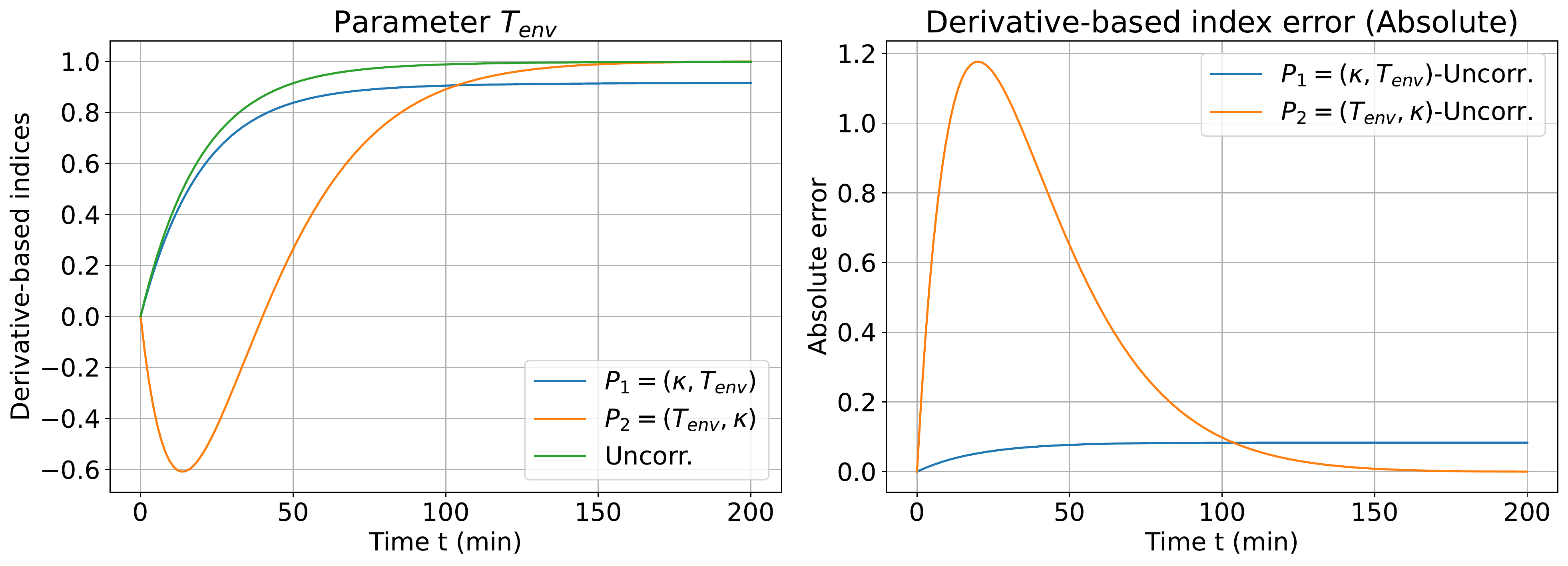}
		\caption{Full ($\mathcal{P}_2$) and Independent ($\mathcal{P}_1$) sensitivity indices of $T_{env}$.}
		\label{fig:coffee_dependent_tenv_der}
	\end{subfigure}
	\caption{Derivative-based indices considering dependent parameters with correlation $\rho_{\mathcal{C}}=0.4$ and the absolute difference with the uncorrelated indices.}
	\label{fig:coffee_dependent_der}
\end{figure}

\begin{figure*}[t!]
	\centering
	\begin{subfigure}[b]{0.7\textwidth}
		\centering
		\includegraphics[width=\textwidth]{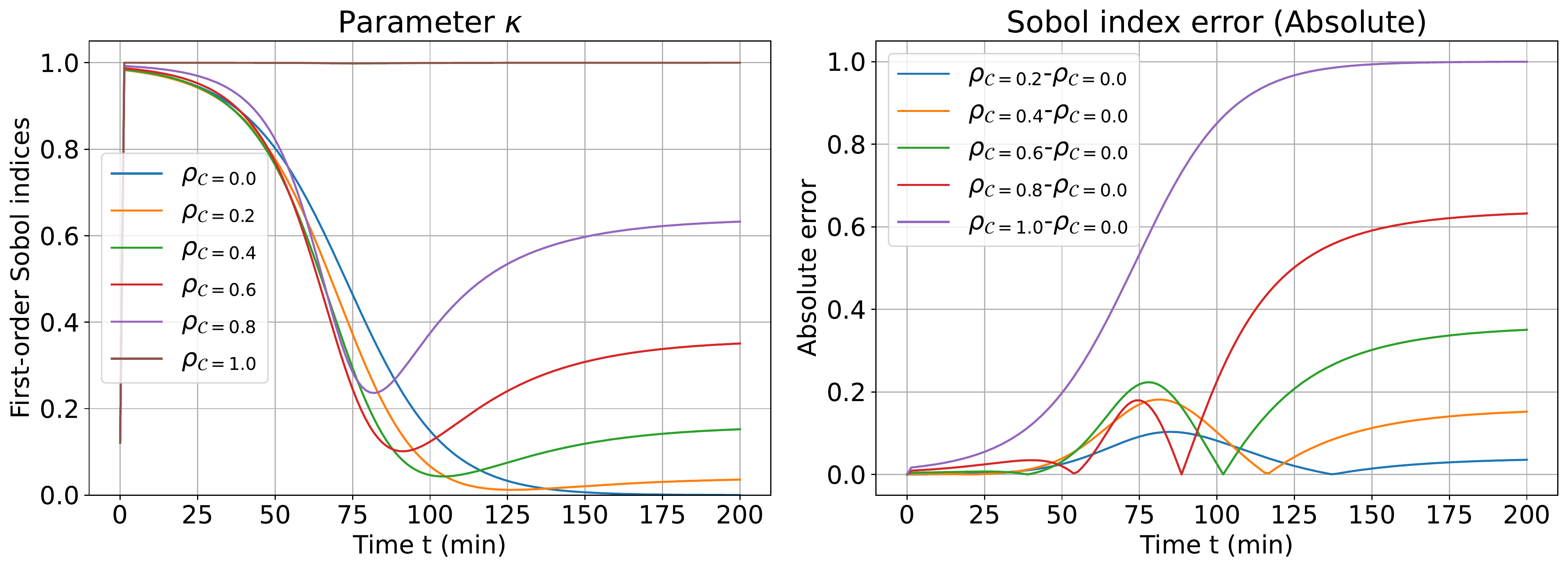}
		\caption{First order  Full index of the $\kappa$ parameter.}
		\label{fig:coffee_dependent_kappa_corr}
	\end{subfigure}
	
	\begin{subfigure}[b]{0.7\textwidth}
		\centering
		\includegraphics[width=\textwidth]{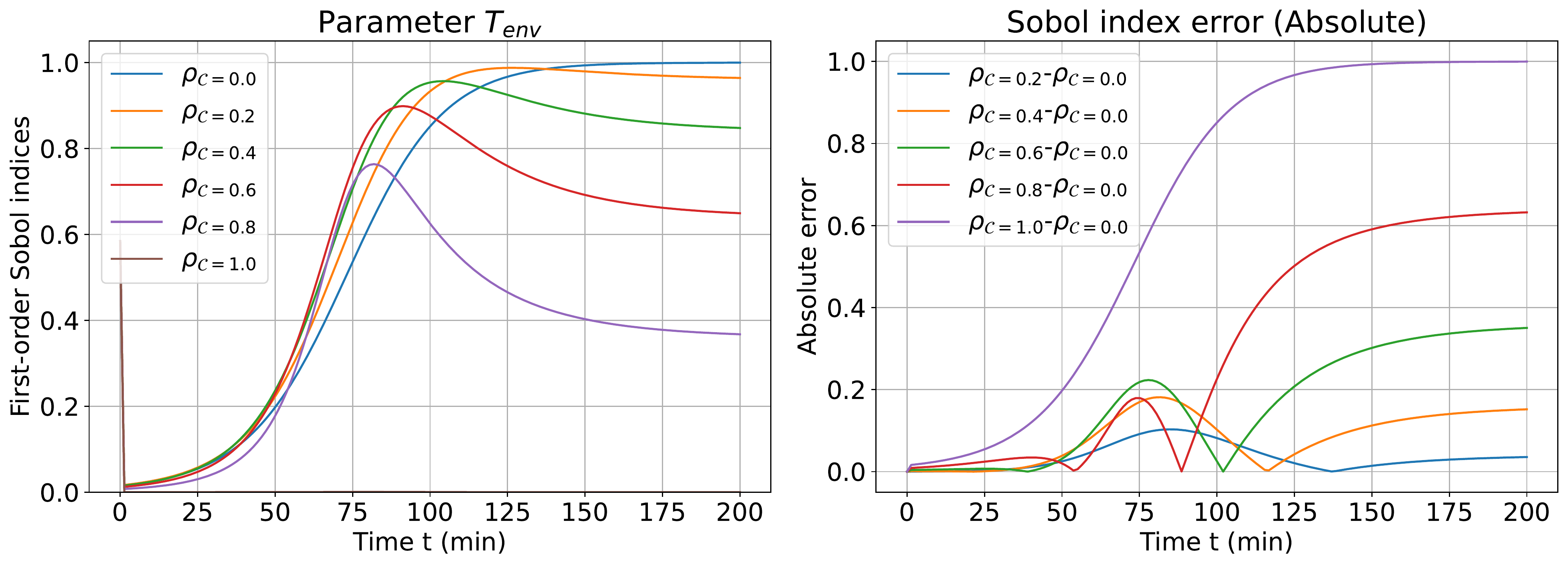}
		\caption{First order Independent index of the $T_{env}$ parameter.}
		\label{fig:coffee_dependent_tenv_corr}
	\end{subfigure}
	\caption{Sobol indices considering dependent parameters with increasing correlation $\rho_{\mathcal{C}}$ and  permutation $(\kappa, T_{env})$, showing also difference with respect to the uncorrelated indices.}
	\label{fig:coffee_dependent_corr}
\end{figure*}

Behavior of the derivative-based indices in the correlated case for the parameter $\kappa$ is shown in Fig.~\ref{fig:coffee_dependent_kappa_der}. In Sec.~\ref{sec:uncorrelated} it was shown that the significance of the heat transfer diminishes toward the end of the simulation time, $t>150\, min$, and the value of the derivative-based index approaches zero. This is due to the fact that the final equilibrium is completely determined by the ambient temperature parameter $T_{env}$. However, when we consider correlation between the parameters and the  Full order index (permutation $\mathcal{P}_1$) the significance of the parameter $\kappa$ is increased since the  Full order index includes also the interaction with the other parameters due to the correlations. In physical terms, it can be interpreted such as when the parameter $\kappa$ is increased, the ambient temperature $T_{env}$ will be increased due to their positive correlation $\rho_{\mathcal{C}} = 0.4$. Consequently, the final temperature of the coffee cup will be increased as well and this is reflected accordingly in the  Full order index of the $\kappa$ parameter which is no longer zero as in the uncorrelated case.
It is also interesting to observe the behavior around the simulation time $t\approx20 \, min$. Note that the magnitude of the  Full order index is reduced compared to the uncorrelated case. The reason for this is again the correlated interaction with the $T_{env}$ parameter. Studying the effect of increasing the  $\kappa$ incurs also an increase in the $T_{env}$ due to their correlation. This however reduces the temperature gradient when assuming constant initial temperature of the coffee cup, thus the cooling process is reduced, even though the heat transfer coefficient was increased. This effect is represented by the reduced magnitude of the  Full order index of $\kappa$ parameter around the simulation time $t\approx20\, min$.

Similar logic applies when considering the Independent index. Consider the Independent index of the $T_{env}$ parameter (permutation $\mathcal{P}_1$) in Fig.~\ref{fig:coffee_dependent_tenv_der}. In the uncorrelated case, the equilibrium near the end of the simulation time, $t>150\, min$, was completely determined by the $T_{env}$ parameter. In the correlated case, after removing the effect of the correlation, the Independent index is proportionally reduced since a part of the ambient temperature growth was induced by the effect of the $\kappa$ parameter and the Independent index eliminates these parameter interactions.

\begin{figure*}[t!]
	\centering
	\begin{subfigure}[b]{0.75\textwidth}
		\centering
		\includegraphics[width=\textwidth]{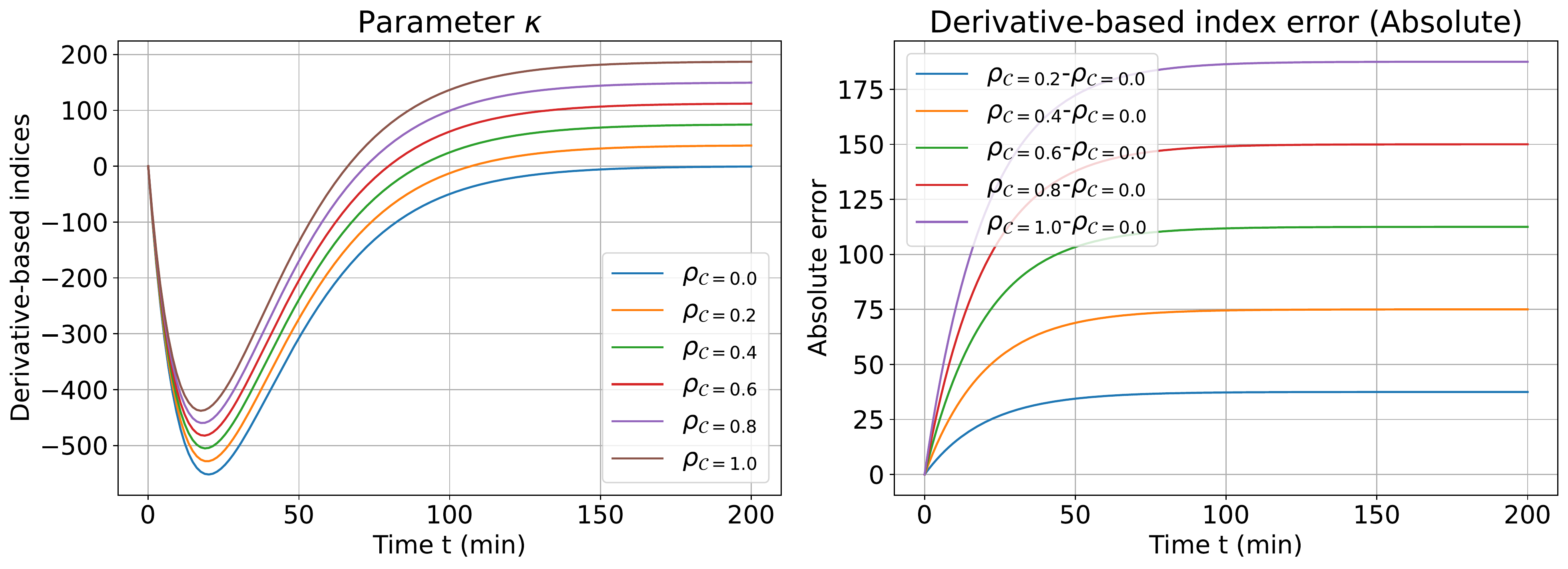}
		\caption{First order Full index of the $\kappa$ parameter.}
		\label{fig:coffee_dependent_der_kappa_corr}
	\end{subfigure}
	
	\begin{subfigure}[b]{0.75\textwidth}
		\centering
		\includegraphics[width=\textwidth]{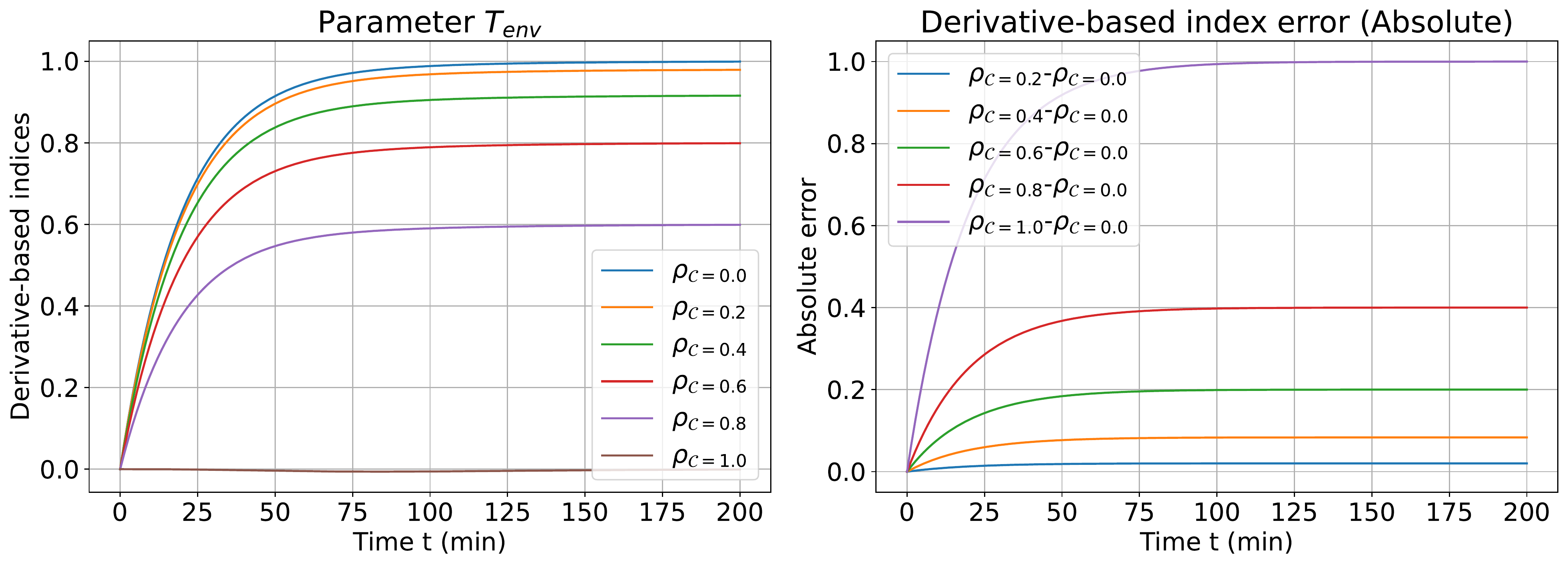}
		\caption{First order Independent index of the $T_{env}$ parameter.}
		\label{fig:coffee_dependent_der_tenv_corr}
	\end{subfigure}
	\caption{Derivative-based indices considering dependent parameters with increasing correlation $\rho_{\mathcal{C}}$ and  permutation $(\kappa, T_{env})$, showing also difference relative to the uncorrelated indices.}
	\label{fig:coffee_dependent_der_corr}
\end{figure*} 


\subsection{Parameters with Increasing Correlation}
It is important to understand the effect of the correlation to the value of the indices.
In this section, the correlation $\rho_\mathcal{C}$ is gradually increasing in increments of $0.2$, ranging from zero all the way to one (i.e. from no correlation up to perfect correlation). The largest value of the correlation is slightly reduced, $\rho_\mathcal{C} = 1.0 - \epsilon, \ \epsilon = 10^{-10}$, in order to preserve positive definiteness of the correlation matrix~$C$.

\subsubsection{Variance-based Sensitivity}
%

%

The study of the First order Sobol indices is first performed considering the permutation $(\kappa, T_{env})$, which is used to compute  Full Sobol index for $\kappa$ and an Independent index for $T_{env}$ shown in Fig.~\ref{fig:coffee_dependent_corr}.
It can be observed that as the correlation increases, the Full index of $\kappa$ at $t>125\, min$ in Fig.~\ref{fig:coffee_dependent_kappa_corr} is gradually increasing, while at the same time the Independent index of the other parameter in Fig.~\ref{fig:coffee_dependent_tenv_corr} proportionally decreases. This reflects the fact that due to the correlation, the Full index $\kappa$ becomes gradually more significant due to its correlation with the $T_{env}$ parameter, not because of the parameter $\kappa$ itself. On the other hand, the amount of the variance explained by $T_{env}$ alone decreases with increasing correlation, because of its interaction with $\kappa$.
In the limit situation when $\rho_{12}=1$, the parameters alone become insignificant and all of the variance is explained by their correlated interaction. Note that the Independent index  is zero (Fig.~\ref{fig:coffee_dependent_tenv_corr})
while the Full index is one (Fig.~\ref{fig:coffee_dependent_kappa_corr}).

Similar effect can be observed using the permutation $(T_{env}, \kappa)$ used to compute Independent Sobol index for $\kappa$ and the Full index for $T_{env}$.

\subsubsection{Derivative-based Sensitivity}

The effect of increasing correlation for the derivative-based sensitivity indices is shown in Fig.~\ref{fig:coffee_dependent_der_corr}. It shows the indices obtained from the permutation  $(\kappa, T_{env})$, which corresponds to the permutation $\mathcal{P}_1$ in Fig.~\ref{fig:coffee_dependent_der}. It can be seen that the increasing correlation intensifies the effects described in Sec.~\ref{sec:parameter_dep}. Note that in the extreme case of correlation $\rho_{\mathcal{C}} = 1.0$ the Independent index becomes zero across the whole simulation, as the parameter $T_{env}$ is completely explained by the parameter $\kappa$. Note that, as opposed to the variance-based sensitivity indices, this doesn't mean that the Full order index is equal to one across the simulation time, since the range of the index values is not bound to the interval $(0,1)$ nor there is any property similar to Eq.~\eqref{eq:sobol_sum}.

\subsection{Convergence Analysis \label{sec:convergence}}
\begin{figure}[t!]
	\centering
	\includegraphics[width=0.49\textwidth]{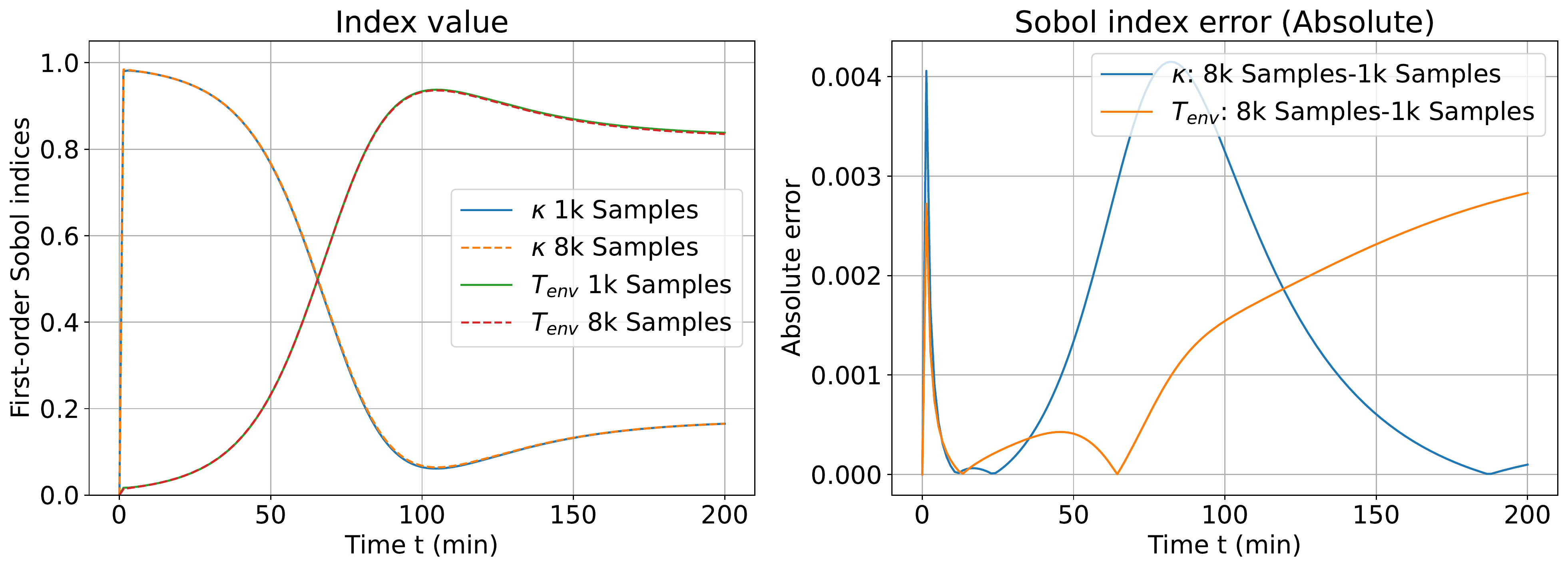}
	\caption{Convergence of the QMC method for Sobol indices using permutation $(\kappa, T_{env})$ with correlation $\rho_{\mathcal{C}}=0.417$.}
	\label{fig:coffee_converg_qmc}
\end{figure}

\begin{figure}[t!]
	\centering
	\begin{subfigure}[b]{0.49\textwidth}
		\centering
		\includegraphics[width=\textwidth]{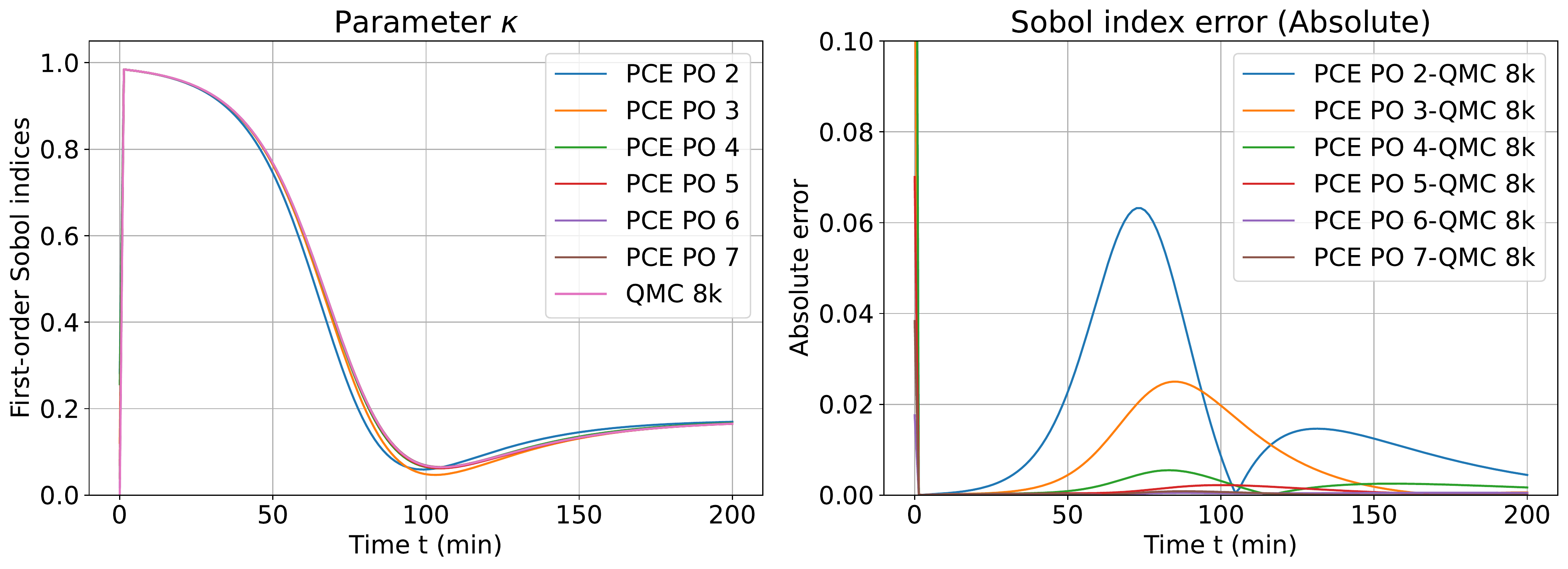}
		\caption{First order index of the $\kappa$ parameter.}
		\label{fig:coffee_converg_kappa_pce}
	\end{subfigure}
	
	\begin{subfigure}[b]{0.49\textwidth}
		\centering
		\includegraphics[width=\textwidth]{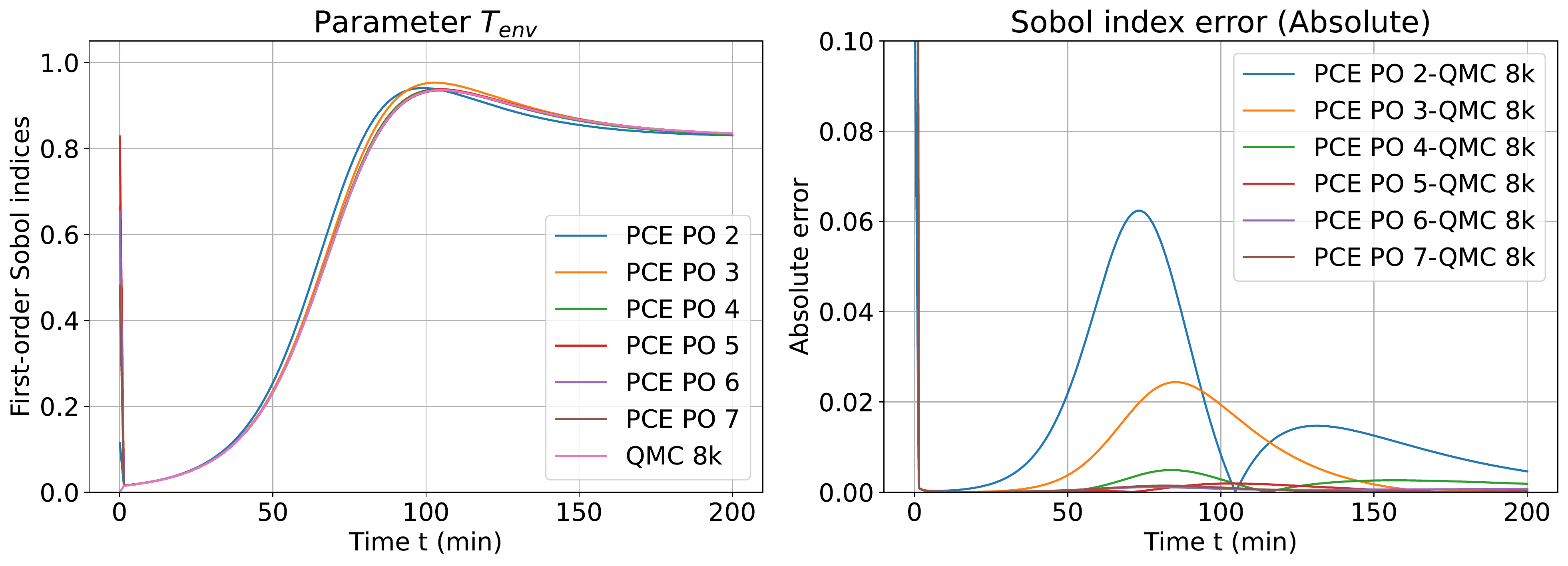}
		\caption{First order index of the $T_{env}$ parameter.}
		\label{fig:coffee_converg_tenv_pce}
	\end{subfigure}
	\caption{Convergence of PCE method for variance-based indices with increasing polynomial order (PO) using permutation $(\kappa, T_{env})$ with correlation $\rho_{\mathcal{C}}=0.417$.}
	\label{fig:coffee_converg_pce}
\end{figure}

\begin{figure}[ht!]
	\centering
	\begin{subfigure}[b]{0.49\textwidth}
		\centering
		\includegraphics[width=\textwidth]{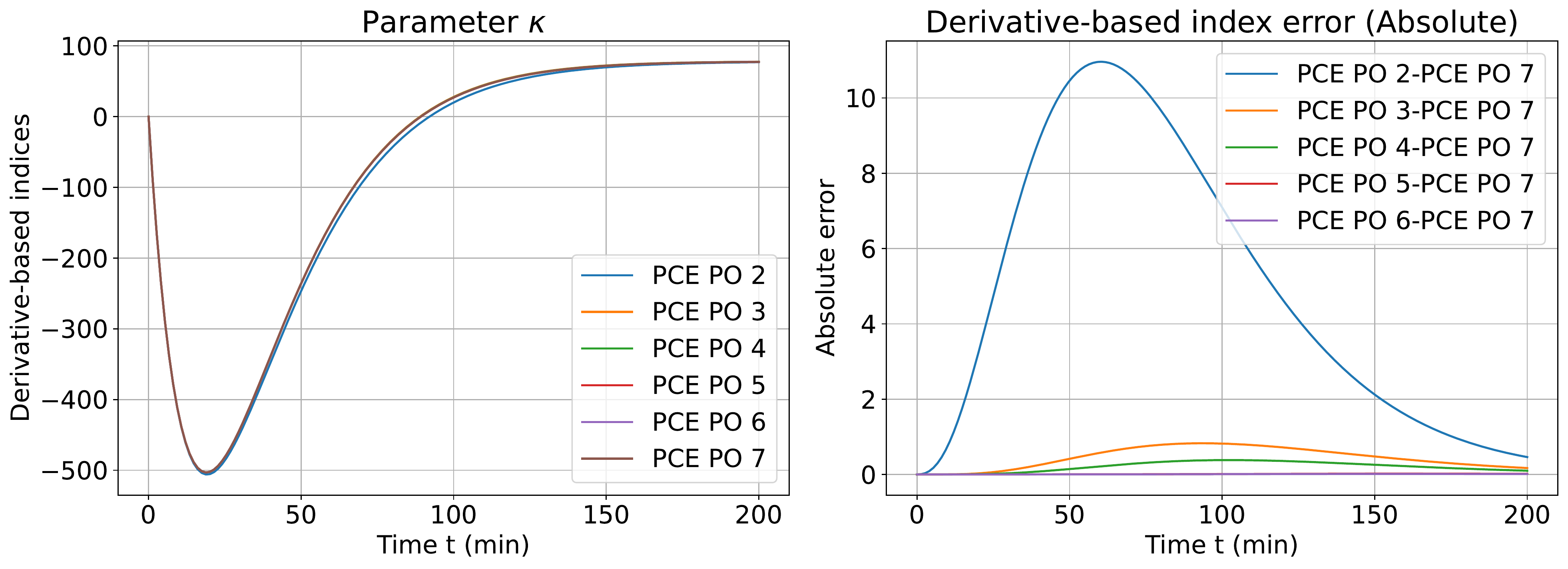}
		\caption{First order index of the $\kappa$ parameter.}
		\label{fig:coffee_converg_der_kappa_pce}
	\end{subfigure}

	\begin{subfigure}[b]{0.49\textwidth}
		\centering
		\includegraphics[width=\textwidth]{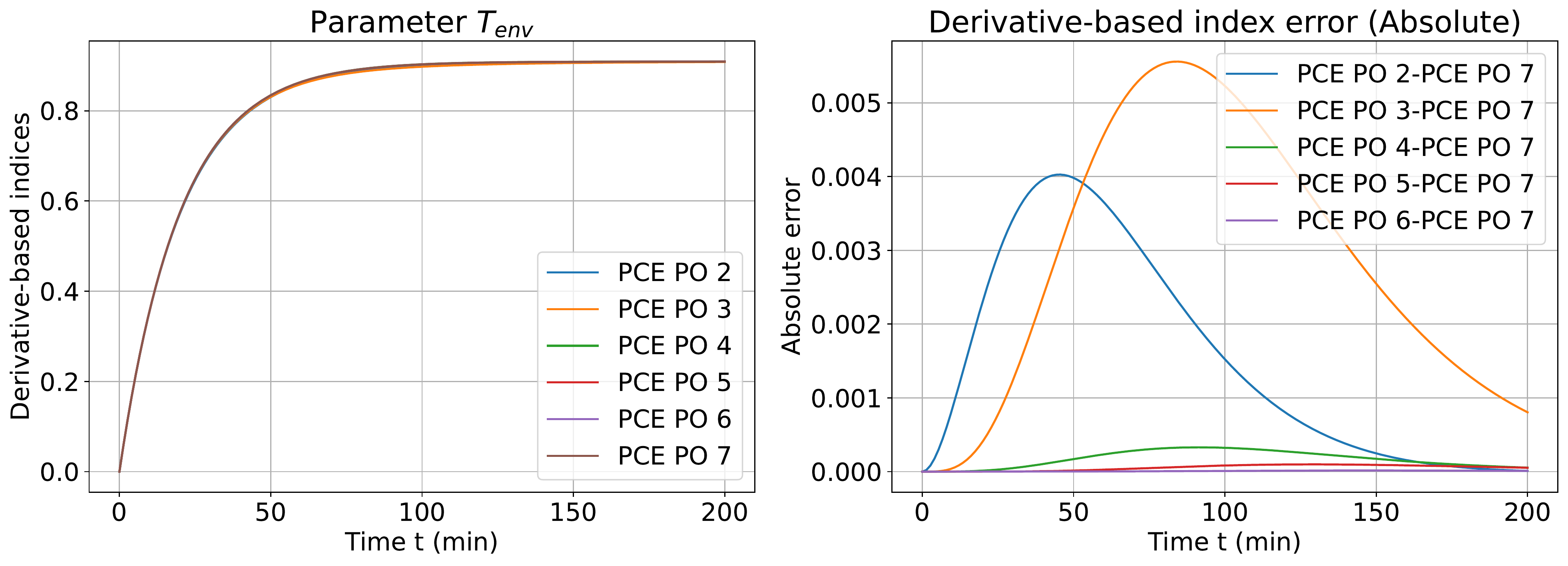}
		\caption{First order index of the $T_{env}$ parameter.}
		\label{fig:coffee_converg_der_tenv_pce}
	\end{subfigure}
	\caption{Convergence of PCE method for derivative-based indices using permutation $(\kappa, T_{env})$ with correlation $\rho_{\mathcal{C}}=0.417$.}
	\label{fig:coffee_converg_der_pce}
\end{figure}

The convergence of the QMC method is tested, which is later used as a reference value for the PCE method. The QMC method is run with an increasing number of samples, and the resulting first order indices are shown in Fig.~\ref{fig:coffee_converg_qmc}. The absolute difference between the indices is well below the threshold of significance $0.05$, thus the method is considered to have converged. The rather arbitrary value of $0.05$ is frequently accepted for this type of analysis for distinguishing important parameters from the unimportant ones~\cite{GSA-cutoff}, thus similar idea can be applied to declare a method to converge.

The PCE method is run with an increasing polynomial order, ranging from 2nd to 7th order. Figure~\ref{fig:coffee_converg_pce} illustrates that the variance-based indices computed by the PCE method differ from the reference value of the QMC by less than $0.05$ for polynomial order three and higher orders. For the fourth polynomial order, the difference  is below $0.01$. For practical purposes the analysis can be run with third of fourth order polynomials. The difference in the initial point is attributed to the fact that the variance in this point is zero and the variance-based Sobol indices are not defined, thus the difference is meaningless here.
Convergence of the derivative-based indices with the surrogate polynomial order are shown in Figure~\ref{fig:coffee_converg_der_pce}.
	\section{Conclusions and Future Work\label{sec:conclusions}}

%

The  SA methods introduced in this paper provides a comprehensive way to quantify the uncertainty and sensitivity of a model with correlated inputs.
As demonstrated in the numerical experiments, the sensitivity indices ignoring parameter correlations significantly differ from their counterparts which do account for the correlation. 
This can have profound implications for assessing the influence of the individual indices on the model sensitivity.
In case of the variance-based sensitivity, the model uncertainty can be over/under-estimated at the presence of the correlation. In case of the derivative-based indices, 
it was demonstrated that the sensitivity of the associated parameters can not only differ in magnitude, but even invert the sign of the derivative-based index, thus reversing the model behavior compared to the prediction of the study disregarding the correlations.
In conclusion, it is essential to consider the parameter correlations during the SA in order to get realistic estimation of the sensitivity indices.


A comprehensible and easy to understand application model in this work was intentionally chosen in order to intuitively understand the input-output interactions and to clearly demonstrate the impact of the input parameter correlations on the sensitivity indices.
In the follow-up work we will apply the method to a large-scale model from the domain of energy markets, where the input parameters such as fossil fuel prices are often correlated. We will evaluate the impact of the correlations on the sensitivity analysis, and explore also associated high-performance computational aspects.

	\bibliography{mybibliography}
	
\end{document}